\documentclass[aps,prd,superscriptaddress,nofootinbib,tighten,preprint]{revtex4-1}

\include{header}
\usepackage[utf8]{inputenc}
\usepackage{amsfonts}
\usepackage{amssymb}
\usepackage{amsmath}
\usepackage{graphicx}
\usepackage{subcaption}
\usepackage{ragged2e}
\usepackage{hyperref}
\usepackage{url}
\usepackage{relsize}
\usepackage{gensymb}
\usepackage[T1]{fontenc}
\usepackage{color}
\usepackage{tabularx}

\usepackage[english]{babel}
\usepackage{lipsum}
\usepackage{float}
\usepackage{lmodern}
\usepackage{ragged2e}
\usepackage{placeins}
\usepackage{etoolbox}
\usepackage{graphicx} % For including figures
\usepackage{caption} % For customizing captions
\usepackage{titletoc}
\usepackage{tikz}
\usepackage{tocbasic}
\usepackage{amsmath}
\usepackage[bottom]{footmisc}
\usepackage{setspace}
\usepackage{multirow}

\hypersetup{
	colorlinks=true,
	linkcolor=blue,
	linktoc=page,
	filecolor=blue,
	urlcolor=blue,
	citecolor=blue,
	bookmarks=true,
	}

\DeclareCaptionJustification{justified}{\justifying}
\titlecontents{section}
[0pt]
{\small\bfseries}
{\thecontentslabel\enspace}
{}
{\titlerule*[0pc]{}\contentspage}

\setcitestyle{numbers,square}

\makeatletter
\def\@hangfrom@section#1#2#3{\@hangfrom{#1#2}#3}%\MakeTextUppercase{#3}}%
\def\@hangfroms@section#1#2{#1#2}%\MakeTextUppercase{#2}}%
\makeatother

\def\slc#1{\setbox0=\hbox{$#1$}           % set a box for #1
    \dimen0=\wd0                                 % and get its size
    \setbox1=\hbox{/} \dimen1=\wd1               % get size of /
    \ifdim\dimen0>\dimen1                        % #1 is bigger
       \rlap{\hbox to \dimen0{\hfil/\hfil}}      % so center / in box
       #1                                        % and print #1
    \else                                        % / is bigger
       \rlap{\hbox to \dimen1{\hfil$#1$\hfil}}   % so center #1
       /                                         % and print /
    \fi}

\newcommand{\beq}{\begin{equation}}
\newcommand{\eeq}{\end{equation}}
\newcommand{\beqa}{\begin{eqnarray}}
\newcommand{\eeqa}{\end{eqnarray}}

\begin{document}
\title{Synergy between DUNE and T2HKK to probe Invisible Neutrino Decay}

\author{\textbf{Zannatun Firdowzy Dey}}
\email{zannatundey@gmail.com}
\affiliation{Assam Don Bosco University, Tapesia Campus, Sonapur, Assam, 782402, India}
%\href{mailto:zannatundey@gmail.com}{zannatundey@gmail.com}

\author{\textbf{Debajyoti Dutta}}
\email{debajyotidutta.hep@gmail.com}
\affiliation{Assam Don Bosco University, Tapesia Campus, Sonapur, Assam, 782402, India}
\affiliation{Bhattadev University, Pathsala Town, Barpeta, Assam 781325}

\bigskip

\begin{abstract}
\noindent
We address the consequence of invisible neutrino decay within the framework of two long base-line neutrino experiments: T2HKK (Tokai-to-Hyper-Kamiokande-to-Korea) and DUNE (Deep Underground Neutrino experiment). Our primary objective is to bring out the aspects of CC (charged current) and NC (neutral current) measurements at DUNE in the context of invisible neutrino decay. We find that the inclusion of NC measurements with the CC measurements enhances its ability to constrain invisible neutrino decay. Further, the synergy between DUNE and T2HKK improves the constraints on invisible neutrino decay. At 3$\sigma$  C.L. (confidence level) the derived constraint is $\tau_{3}/m_{3}\geq6.21\times10^{-11}$ s/eV.  Additionally, if nature prefers $\nu_{3}$ to be unstable and the decay width is $\tau_{3}/m_{3}= 2.2\times10^{-11}$ s/eV, this combination can exclude the no-decay scenario at more than 5$\sigma$ C.L. Although the CP sensitivity is not much hindered in the presence of invisible neutrino decay, the measurements of $\theta_{23}$ and the ability to resolve octant of $\theta_{23}$ is significantly influenced in these individual experiments. In the presence of invisible neutrino decay, the synergy between DUNE and T2HKK can exclude the wrong octant somewhat more effectively than either experiment alone.

\end{abstract}
\newpage

\maketitle
\hrulefill
\tableofcontents
\hrulefill

\section{Introduction}
The discovery of neutrino oscillations has led to a new area of research Beyond the Standard Model (BSM), explaining the existence of neutrino mass. Out of the six oscillation parameters, $\theta_{12}, \theta_{13}$ and $\Delta{m}^2_{21}$ are measured precisely. However, the sign of $\Delta{m}^2_{31}$, octant, absolute value of $\theta_{23}$  and the CP violating phase $\delta_{\mathrm{CP}}$ are yet to be measured precisely. There are few indications regarding these unknown neutrino oscillation parameters from the ongoing neutrino oscillation experiments T2K \citep{T2K:2017hed} and NO$\nu$A \citep{NOvA:2016kwd}. Both the experiments predict the same best-fit value for $\sin^2\theta_{23}$ at 1$\sigma$ C.L. Yet they have strong disagreements on the measured best-fit value of $\delta_{\mathrm{CP}}$. With more data from these experiments or from a combined analysis of the present data, better constraints can be put on these oscillation parameters. In these contexts, the upcoming experiments like DUNE\citep{DUNE:2020lwj,DUNE:2020ypp,DUNE:2020mra,DUNE:2020txw, DUNE:2020fgq, DUNE:2022aul,DUNE:2020jqi}, T2HK/T2HKK\cite{Hyper-KamiokandeProto-:2015xww, hyper2018physics}, ESS$\nu$SB\cite{ESSnuSB:2013dql}, JUNO\cite{JUNO:2015zny}, INO\cite{ICAL:2015stm}, PINGU\cite{IceCube-PINGU:2014okk}, KM3Net-ORCA\cite{deSalas:2018kri} etc. are aiming to determine these unknown parameters with more precision. %Since the discovery of neutrino oscillation has made huge progress and this phenomenon is now well established, a complete understanding of neutrino oscillations will provide an opportunity to explore new physics beyond the standard model.% 
In the present scenario, BSM physics such as the existence of sterile neutrinos, vector and scalar non-standard interactions (NSIs), invisible neutrino decay, etc., are some interesting topics to be explored further, as the presence of such new physics will affect the measurements of the standard neutrino oscillation parameters in these upcoming experiments. These long and short-baseline neutrino experiments will be able to address some of the questions.

Invisible neutrino decay is a new topic of interest, which was first proposed by ref. \cite{Bahcall:1972my} to explain the solar neutrino problem. There are two possible modes of neutrino decay: visible and invisible decay. Both Dirac and Majorana neutrinos can undergo invisible decay. For Dirac neutrinos, the coupling between the neutrinos and the light scalar boson (s) leads to the decay mode: $\nu_{j}\rightarrow\bar{\nu}_{iR}+\chi$, where $\bar{\nu}_{iR}$ is the right-handed singlet and $\chi$ is the iso-singlet scalar. If neutrinos are Majorana particles, then the decay mode is $\nu_{j}\rightarrow\nu_{s}+J$, where $\nu_{s}$ is the sterile neutrino and $J$ is a Majoron coupled by a pseudo-scalar coupling. This Majoron has to be singlet as constrained by the LEP data \citep{Pakvasa:1999ta}. In both scenarios, the decay products are sterile and are invisible to the detectors. If the neutrino decays via $\nu_{j}\rightarrow\nu_{i}+J$ and $\nu_{j}\rightarrow \bar{\nu}_{i}+J$ channels, then the final states can be detected, and these modes of decay are known as visible neutrino decay. In visible neutrino decay, the final state fermion is a lighter neutrino, which is visible to the detectors, while in the other case, the final state particle, being a sterile neutrino, cannot be detected.

Assuming $\nu_{2}$ to be unstable, neutrino decay with oscillation has been studied in \cite{Acker:1993sz, Choubey:2000an, Bandyopadhyay:2001ct, Joshipura:2002fb}. These studies put constraints on the lifetime of $\nu_{2}$. In \citep{Bandyopadhyay:2002qg} solar neutrino data provide a constraint at $\tau_{2}/m_2 > 8.5 \times 10^{-7}$ s/eV. Constraints on both $\tau_{2}$ and $\tau_{1}$ have also been derived from low-energy solar neutrino data and supernova neutrino observations \citep{Berryman:2014qha}. The lifetime of $\nu_e$ is strongly constrained by SN1987A data \citep{Frieman:1987as}. %Constraints on $\nu_3$ lifetime come from extensive studies in atmospheric and long-baseline neutrino experiments.%
Earlier, to overcome the atmospheric neutrino problem, the possibility of neutrino decay was proposed in \cite{LoSecco:1998cd} but the fit was very poor. In \cite{Barger:1998xk, Lipari:1999vh}, authors studied
neutrino decay with mixing to fit Super-Kamiokande (SK) data. Later, in \cite{Super-Kamiokande:2004orf} the SK collaboration showed that its data could better fit neutrino oscillation only. However, scenarios where an unstable neutrino decays to a sterile neutrino with $\Delta m^2\sim 0.003$ eV$^2$ show potential for improving the fit to SK data \cite{Choubey:1999ir}. So decay with oscillation can still give a better fit to SK data. Global analyses combining MINOS and atmospheric data have placed constraints such as  $\tau_{3}/m_3 \geq 2.9 \times 10^{-10}$ s/eV at 90$\%$ C.L \cite{Gonzalez-Garcia:2008mgl}. Bounds on ${\tau_3}/{m_3}$ considering two generations of neutrinos have been calculated using MINOS and T2K data in \cite{Gomes:2014yua}. Using the preliminary data, the  T2K and NO$\nu$A experiments provide further insights, indicating  ${\tau_3}/{m_3} > 1.5 \times 10^{-12}$ s/eV at 3$\sigma$ C.L. \cite{Choubey:2018cfz}.  In \citep{Ternes:2024qui}, authors have presented constraints using data from NO$\nu$A, T2K, MINOS/MINOS+, and their combinations assuming NO(IO) and at $90\%$ C.L. the bounds are $3.0(2.9)\times10^{-12}$ s/eV, $9.9(7.0)\times10^{-12}$ s/eV, $1.6(1.8)\times10^{-11}$ s/eV and $2.4(2.4)\times10^{-11}$ s/eV respectively. Authors in \cite{Denton:2018aml} have also explored the implications of invisible neutrino decay, such as its potential to resolve discrepancies between track and cascade observations at IceCUBE. Authors of \citep{Gronroos:2024jbs} have done an interesting analysis of the oscillation probabilities in the presence of invisible neutrino decay using the well-known Cayley-Hamilton formalism.

Numerous studies in the literature have explored the prospective capabilities of forthcoming experiments in detecting invisible decay. In the context of invisible neutrino decay, the performance of DUNE has been studied in \cite{Choubey:2017dyu}. A runtime of 5+5 years at DUNE is expected to yield constraints of ${\tau_3}/{m_3}>4.50\times10^{-11}$ s/eV at $90\%$ C.L. Incorporating NC measurements with CC measurements at DUNE is anticipated to further enhance this constraint \cite{Ghoshal:2020hyo}. Additionally, the MOMENT experiment is projected to achieve bounds of ${\tau_3}/m_3$ $>1.6\times10^{-11}$ s/eV at $3\sigma$ C.L. \cite{Tang:2018rer}. The ESSnuSB experiment, as discussed in \citep{Choubey:2020dhw}, demonstrates significant potential with 3$\sigma$ bounds on $\tau_3$ at different baselines. Specifically, at 540 km, the constraints are less stringent compared to the 360 km baseline. Furthermore, the combined sensitivity of T2HK+ESS$\nu$SB is expected to constrain ${\tau_3}/{m_3} \geq 4.36 \times 10^{-11}$ s/eV at 3$\sigma$ C.L., while T2HKK+ESS$\nu$SB could achieve ${\tau_3}/{m_3} \geq 5.53 \times 10^{-11}$ s/eV at the same confidence level \cite{Chakraborty:2020cfu} and at 3$\sigma$ C.L. The JUNO experiment, detailed in \citep{Abrahao:2015rba}, expects to set a stringent constraint of ${\tau_3}/{m_3} > 7.5 \times 10^{-11}$ s/eV at $95\%$ C.L. with 100 kt.years exposure. With a total exposure of 70 ton.year, solar data using high-resolution multi-ton Xenon detectors can constrain $\tau_{2}/m_{2}$ and  $\tau_{1}/m_{1}$ at 2$\sigma$ C.L. and the bounds are $\tau_{2}/m_{2} > 8 \times 10^{-3}$ s/eV and $\tau_{1}/m_{1} > 3 \times 10^{-2}$ s/eV, respectively \citep{Huang:2018nxj}. If $\alpha_{i} = m_{i}/\tau_{i}$ is the decay parameter (where index $i$ indicates mass eigenstate) then 10 years of solar data may provide stringent bounds on $\alpha_{1}$ and $\alpha_{2}$ and at 2$\sigma$ C.L., the bounds are $\alpha_{1} < 7.8 \times 10^{-15}$ eV$^2$ and $\alpha_{2} < 2.3 \times 10^{-14}$ eV$^2$ respectively \citep{Martinez-Mirave:2024hfd}. Visible neutrino decay is also tightly constrained, with analyses conducted in the context of long-baseline experiments \cite{Gago:2017zzy, Coloma:2017zpg, Ascencio-Sosa:2018lbk}.

Neutrino physics is in its precision era and upcoming neutrino experiments strive to achieve precise measurements of neutrino oscillation parameters. However, the existence of parameter degeneracy \citep{Barger:2001yr} among the mixing parameters introduces challenges in accurately determining these parameters. Amalgamating different neutrino experiments helps in resolving parameter degeneracy. %In the context of invisible neutrino decay, a combined analysis of T2HK plus ESS$\nu$EB and T2HKK plus ESS$\nu$EB has been performed in \cite{Chakraborty:2020cfu} and at 3$\sigma$ C.L. these combinations can constrain ${\tau_3}/{m_3}$ as ${\tau_3}/{m_3}\geq 4.36\times {10}^{-11}$ s/eV and ${\tau_3}/{m_3}\geq 5.53\times {10}^{-11}$ s/eV respectively. The bounds obtained for the combinations of experiments are better than the individual bounds of these experiments calculated in the paper.% 
In this paper, we perform a synergistic study of DUNE and T2HKK in the context of invisible neutrino decay. Both DUNE and T2HKK are two promising upcoming long-baseline neutrino experiments. DUNE is a beam-based high-statistics experiment with a large detector mass, long baseline, and sufficient matter effect. On the other hand, T2HKK is also a very special experiment where it has two detector sites, and the detectors are placed close to their 1st and 2nd oscillation maxima. It also has a huge detector mass, huge statistics, and sufficient sensitivity to the matter effect, like DUNE. So combining such long-baseline experiments enhances overall sensitivity and aids in identifying potential synergies among the experiments. 

Although separate studies have been performed in the presence of invisible neutrino decay at DUNE and T2HKK, we perform a combined study of these two experiments to explore possible synergies among them. Regarding DUNE, we individually investigate the sensitivity to ${\tau_3}/{m_3}$ for both CC and NC measurements, and then combine the NC with CC. With the combined CC and NC measurements at DUNE, we incorporate data from the T2HKK experiment to get further enhanced limits on ${\tau_3}/{m_3}$. Even while NC events can be measured in water quite effectively, the migration matrix that reconstructs the neutrino events in T2HKK is not available to us. We have not performed NC analysis for T2HKK since it is inappropriate to use a Gaussian energy resolution function in this situation. The analysis is performed within the framework of three-flavour neutrino oscillation scenarios in the presence of matter effect. Additionally, we extensively study the correlations between $\theta_{23}$ and ${\tau_3}/{m_3}$. We thoroughly investigate how the presence of invisible neutrino decay, if it occurs naturally, affects the octant of $\theta_{23}$ and the CP violation sensitivities at the far detectors of DUNE and T2HKK. The same is also explored for the combination of the two experiments.
 
This paper is structured as follows: We address the theory of invisible neutrino decay in Section \ref{sec: theory}. In Section \ref{sec: exp} we give a brief description of the two experiments and the simulation details. Our results are presented in Section \ref{sec: result} and we conclude in Section \ref{sec: summ}.

\section{Theory of invisible neutrino decay}
\label{sec: theory}
We suppose that the heaviest mass eigenstate $\nu_3$ is not stable and undergoes decay into a sterile state, $\nu_4$, with lifetime $\tau_3$ into a new sterile state and a scalar singlet $J$; $(\nu_3\rightarrow\bar{\nu_4}+J)$. Additionally, we assume that there is no mixing of the three active neutrinos with the sterile neutrino. Taking this into account, the neutrino mixing matrix can be described using the standard three-family mixing matrix $U$ in the following manner \\

\begin{equation}
\begin{pmatrix}
\nu_\alpha\\
\nu_s
\end{pmatrix}=
\begin{pmatrix}
U & 0\\0 & 1
\end{pmatrix}
\begin{pmatrix} \nu_i \\ \nu_4
\end{pmatrix}
\label{eq: sterile matrix}
\end{equation}

Where $U$ is the standard PMNS (Pontecorvo–Maki–Nakagawa–Sakata matrix) matrix describing the standard three neutrino oscillation, $\alpha\rightarrow\nu_e,\nu_\mu,\nu_\tau$ indicates the flavour eigenstate and $i=1,2,3$ the mass eigenstates. We assume normal mass ordering ($m_3>m_1$) throughout this work. Furthermore, we assume that the mass and decay eigenstates are identical. Based on these presumptions, the evolution of neutrinos in the presence of matter can be expressed as follows:

\begin{equation}
{i\frac{d}{dx}}{\begin{pmatrix}
\nu_{e}\\
\nu_{\mu}\\
\nu_{\tau}
\end{pmatrix}}=\left[ U \left[ \frac{1}{2E}{\begin{pmatrix}
0 & 0 & 0\\
0 & \Delta{m}^2_{21} & 0\\
0 & 0 & \Delta{m}^2_{31}
\end{pmatrix}}-{i\frac{m_3}{2E{\tau_3}}}{\begin{pmatrix}
0 & 0 & 0\\
0 & 0 & 0\\ 
0 & 0 & 1
\end{pmatrix}}\right] {U^\dagger} + {\begin{pmatrix}
A_{cc} & 0 & 0\\
0 & 0 & 0 \\
0 & 0 & 0
\end{pmatrix}}\right]{\begin{pmatrix}
\nu_{e}\\
\nu_{\mu}\\
\nu_{\tau}
\end{pmatrix}}
\label{eq: 3matrix}
\end{equation}
Simplifying the equation \ref{eq: 3matrix} we can write,

\begin{equation}
i\frac{d}{dx}v_f=\frac{1}{2E}\left[U\widetilde{H}{U^\dagger}+A\right]{v_f}.
\end{equation}

Where,
\begin{equation}
\widetilde{H}=
\begin{pmatrix} 
0 & 0 & 0 \\
0 & \Delta{m}^2_{21}& 0\\
0 & 0 & \Delta{m}^2_{31}-\frac{{i}m_3}{\tau_3} \\
\end{pmatrix},  
 A=\begin{pmatrix} 
A_{cc} & 0 & 0 \\
0 & 0 & 0\\
0 & 0 & 0 \\
\end{pmatrix}
\label{eq: hamiltonian}
\end{equation}

Here $A_{cc}=2\sqrt{2}G_Fn_eE$ represents the matter potential due to neutrino electron scattering, $G_F$ is the Fermi coupling constant, $E$ is the energy and $n_e$ is the density of the electron. The probability of getting a neutrino from an initial state $\nu_a$ to a final state $\nu_b$ can be expressed as 
\begin{equation}P_{ab}= {|<\nu_b|\nu_a>|}^2.\end{equation}

Where, $a$, and $b$ correspond the flavour states $e, \mu,$ and $\tau$. The decay rate of the unstable state ${\nu_3}$ is defined as ${m_3}/{\tau_3}$. Therefore, in the probability equation, the effect of the decay appears as exp $[-({m_3}/{\tau_3})\frac{L}{E}]$. Hence, a lower value of the decay parameter will affect an experiment with a longer baseline or lower energy.

\section{ Experimental and simulation details}
\label{sec: exp}
\noindent
DUNE and T2HKK are two long baseline experiments. In this section, we give a brief description of these two experiments.

\subsection{DUNE}
DUNE (Deep Underground Neutrino Experiment)\citep{DUNE:2020lwj,DUNE:2020ypp,DUNE:2020mra,DUNE:2020txw, DUNE:2020fgq, DUNE:2022aul,DUNE:2020jqi} plans to shoot a special beam of neutrinos (and their anti-particles) from Fermilab in Illinois. A 120 GeV proton beam with a beam power of 1.2 MW, will hit the graphite target each year to produce around $1.1\times10^{21}$ protons. As a result, charged mesons will be produced, which will decay in flight into neutrinos. The underground far detector will be installed in South Dakota, 1285 km away from the source at Fermilab. The net volume of the Liquid Argon (LAr) detector at the far site will be 40 kt, and in this analysis, we assume a single-phase detector \citep{DUNE:2020ypp}. The whole experiment is set to run for 10 years, spending equal time on neutrinos and anti-neutrinos. The key scientific objectives of the DUNE experiment involve measuring the CP phase, determining neutrino mass ordering, identifying the mixing angle $\theta_{23}$ and its octant, and examining the three-neutrino paradigm. A significant focus is on the search for CP violation in neutrino oscillations, which could provide insights into the matter-antimatter asymmetry. DUNE also endeavours to explore the phenomenon of proton decay. The experiment seeks to enhance its capability to detect neutrinos emitted from core-collapse supernovae and to explore phenomena beyond the scope of the SM \citep{DUNE:2020ypp}.\\
 Our results are obtained using the DUNE simulation configuration given in ref. \citep{DUNE:2021cuw}. For the analysis of CC, we have used the 0.125 GeV energy binning approach with reconstructed energies \(E_{\text{rec}}\) spanning from 0 to 110 GeV. This range is divided into 64 bins spanning from 0 to 8 GeV, each bin being 0.125 GeV wide. The remaining 16 bins cover the range from 8 to 110 GeV, with varying widths.  For the analysis of NC, we have used the migration matrices from ref. \citep{DeRomeri:2016qwo} and the bin size of the used migration matrices is 50 MeV. Therefore in this work, we have also used energy bins of 50 MeV for NC events. We assumed a detection efficiency of $90\%$ for NC events.

\subsection{T2HKK}

The next Hyper-K experiment is planned to construct two identical water-Cherenkov detectors, each 187 kt in size. The first will be located at the Tochibora mine in Japan, with a baseline of 295 km from the J-PARC neutrino target and an off-axis angle of 2.5$^\circ$. The other detector may be located in Korea at a baseline of 1100 km. The flux at 295 km will cover the first oscillation maximum and the flux at  1100  km will cover the second oscillation maximum. The first and second oscillation maxima of the PMNS neutrino appearance probability can be reached due to Korea's longer baseline. The longer baseline of the Korean sites improves the oscillation probability's CP-violating component and resolves parameter combinations between the CP-violating phase and neutrino mass ordering that would be almost degenerate if measurements were limited to the Japanese site.T2HKK offers a baseline that is nearly as long as the proposed DUNE experiment, but it is in an energy band that is similar to that of the current T2K experiment. This enables it to probe oscillations at the second oscillation maximum, a capability that is only shared by the proposed ESS neutrino beam and not available to any other experiment\cite{Hyper-Kamiokande:2016srs}.

\newcolumntype{P}[1]{>{\centering\arraybackslash}p{3cm}}
\newcolumntype{P}[1]{>{\centering\arraybackslash}p{4cm}}
\begin{table}[htbp]
\renewcommand{\arraystretch}{2}
\begin{center}

\begin{tabular}{|P{5cm}|P{5cm}|P{5cm}|}
\hline
 Parameter & Best-fit Value & $3\sigma$ range \\
\hline 
 $\theta_{13}$  & $8.6^\circ$  & [$8.1^\circ, 8.9^\circ$] \\
\hline
 $\theta_{12}$  & $33.82^{\circ}$  & [$31.3^\circ, 35.9^\circ$]  \\
\hline
 $\theta_{23}$  & $41.95^{\circ}$ (LO), $48.44^\circ$ (HO)  & [$38^\circ, 52^\circ$] \\
\hline
 $\Delta m^2_{21}$ (eV$^2$)  & $7.39\times 10^{-5}$  & [$6.82, 8.04] \times 10^{-5}$\\
\hline
$\Delta m^2_{31}$ (eV$^2$)  & $2.52\times 10^{-3}$  & [$2.43, 2.60]\times10^{-3}$ \\
\hline
$\delta_{\mathrm{CP}}$  & $-90^\circ$  & [${-180}^{\circ}, 180^{\circ}$] \\
\hline

\end{tabular}
\caption{The best-fit values of the neutrino oscillation parameters and the corresponding 3$\sigma$ allowed ranges.  }
\label{tab: 3sigma}
\end{center}
\end{table}

\begin{table}[htbp]
\renewcommand{\arraystretch}{2}
\begin{center}
\scriptsize
\begin{tabular}{|l|p{1cm}|p{1cm}|p{1.5cm}|p{1.5cm}|p{1cm}|p{1.5cm}|p{1.5cm}|p{1.5cm}|p{1.5cm}|p{1cm}|p{1.5cm}|}\hline
Experiment & \multicolumn{6}{c|}{Signal} & \multicolumn{5}{c|}{Background}\\
           & App. $\nu$  & App. $\bar\nu$ & Disapp. $\nu$ & Disapp. $\bar\nu$ & NC & tilt & $\nu_{e},\bar\nu_{e}$ CC & $\nu_{\mu},\bar\nu_{\mu}$ CC & $\nu_{\tau},\bar\nu_{\tau}$ CC & NC & tilt \\\hline
DUNE   & $2\%$ & $2\%$ & $5\%$ & $5\%$ & $5\%$ & $5\%$ & $5\%$ & $5\%$ & $20\%$ & $10\%$ & $10\%$ \\
T2HKK  & $3.2\%$ & $3.9\%$ & $3.6\%$ & $3.6\%$ & - & $5\%$ & $5\%$ & $5\%$ & - & - & $10\%$ \\
\hline
\end{tabular}
\caption{signal and background uncertainties associated with DUNE and T2HKK.}
\label{tab: taberror}
\end{center}
\end{table}

In our analysis, we have used GLoBES (General Long Baseline Neutrino Experiment simulator) \citep{Huber:2004ka} software package to simulate the DUNE and T2HKK experiments. We have simulated DUNE for a run time of 10 years, which is equally divided between neutrinos and anti-neutrinos. Similarly, for T2HKK we have considered a run time of 2.5 years for neutrinos and 7.5 years for anti-neutrinos, respectively.  For the energy resolution of T2HKK, we have assumed a Gaussian function of width $15\%/\sqrt{E}$. Table \ref{tab: 3sigma} provides the values of the three neutrino oscillation parameters that we used in our simulation. The values provided in the table align with the current global fit ranges \citep{Esteban:2020cvm}. The calculations have been done considering normal mass ordering. The signal and background normalization uncertainties of DUNE and T2HKK experiments for various channels used in this analysis are given in Table \ref{tab: taberror}.

\section{Results}
\label{sec: result}

In this section, we present the results of our analysis.

\subsection{Probability at DUNE and T2HKK}
\begin{figure}[htbp]
\begin{center}
\includegraphics[scale=0.48]{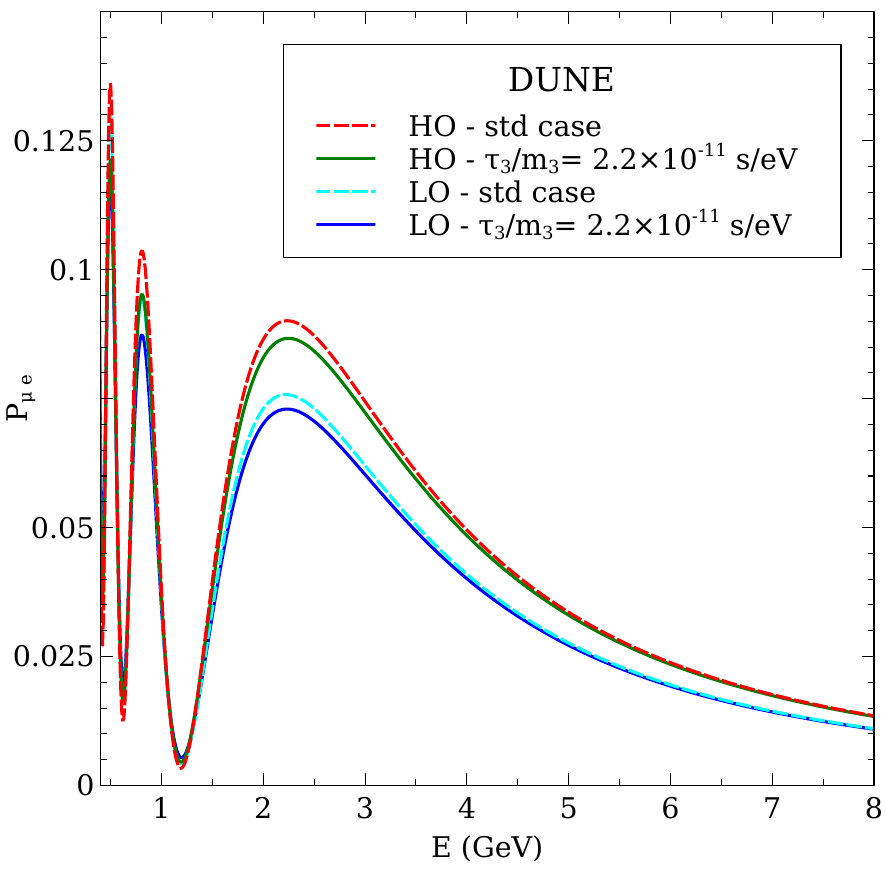} 
\includegraphics[scale=0.48]{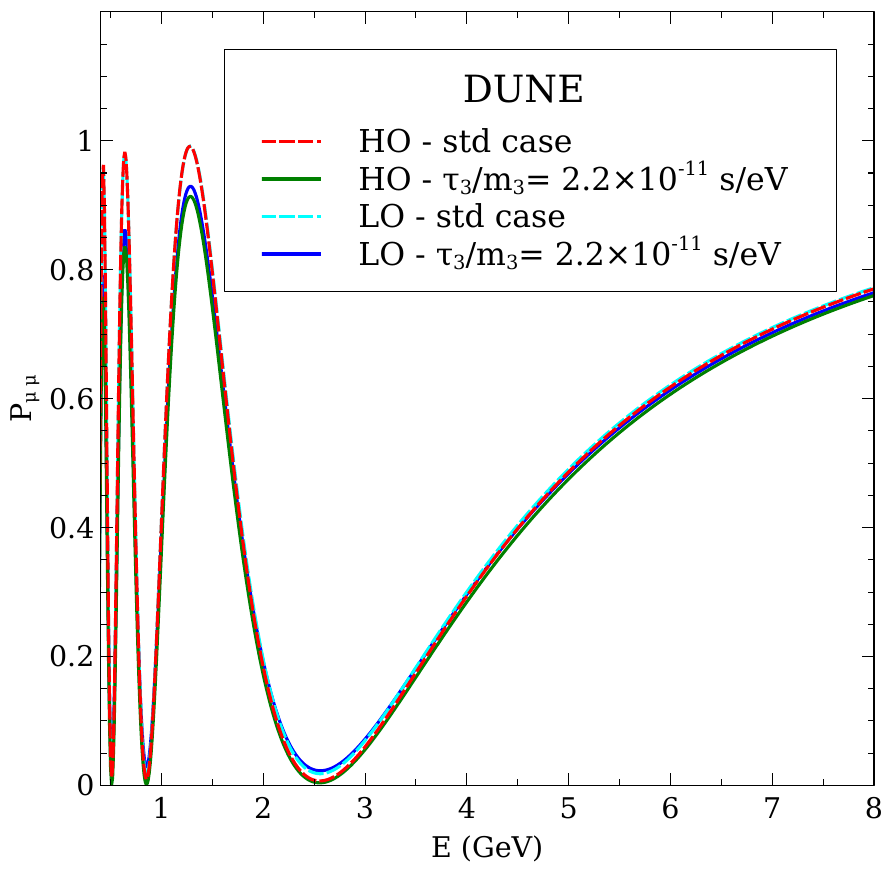} 
\includegraphics[scale=0.48]{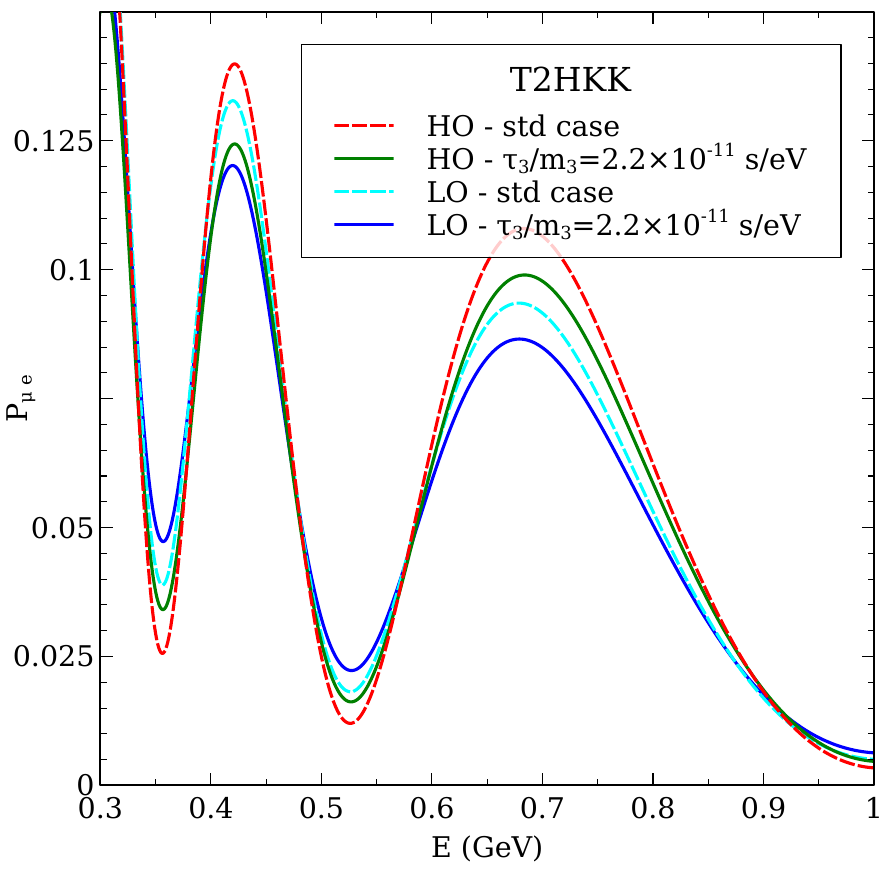} 
\includegraphics[scale=0.48]{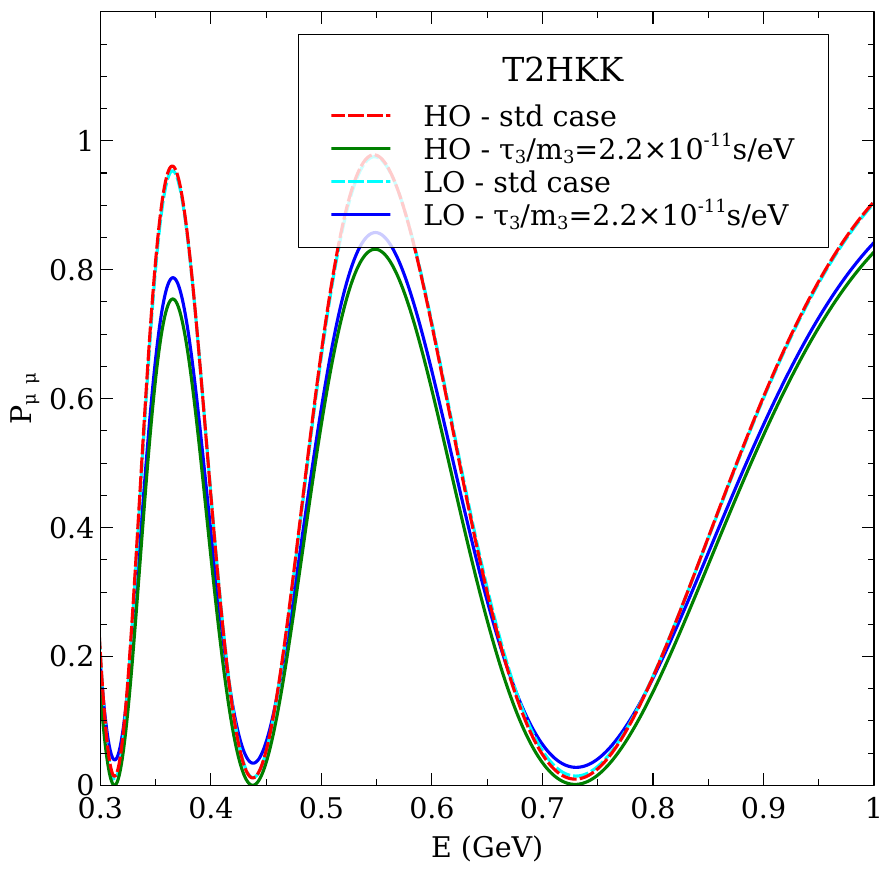}

\caption{The appearance (left panels) and disappearance (right panels) channels of neutrino oscillation probabilities for DUNE and T2HKK experiments as a function of neutrino energy. The top panel corresponds to DUNE, and the bottom panel corresponds to T2HKK.} 
\label{fig: prob}
\end{center}
\end{figure}

In Figure \ref{fig: prob}, we present the probability of neutrino oscillations for the DUNE and T2HKK experiments as a function of energy. The figure illustrates probabilities for both appearance (left) and disappearance (right) channels, comparing scenarios with and without invisible neutrino decay. In the top panel, we present the probability plots for the DUNE and the bottom panel shows the probability plots for T2HKK. For all the cases, the dashed lines are for standard oscillation case and the solid lines are for ${\tau_3}/{m_3}=2.2\times 10^{-11}$ s/eV. We show the results for both the lower octant ($\theta_{23}=41.95^{\circ}$) and the higher octant ($\theta_{23}=48.44^{\circ}$). The red (cyan) dashed lines represent the probability of no decay in HO (LO), while the green (blue) solid lines represent probability plots in the presence of invisible neutrino decay in HO (LO).

From Figure \ref{fig: prob}, we observe that $P_{\mu e}$ decreases in the presence of neutrino decay both in the lower and the higher octants. For a fixed value of ${\tau_3}/{m_3}=2.2\times 10^{-11}$ s/eV the probability in the lower octant is smaller than the probability in the higher octant. Also around the oscillation maxima, the effect of decay is more prominent in T2HKK than DUNE. The disappearance probabilities ($\nu_{\mu}\rightarrow \nu_{\mu}$) are shown in the right panel of Figure \ref{fig: prob}. The disappearance probability plots for the no decay case overlap in both the lower and the higher octants and this behavior is the same in both DUNE and T2HKK. But in the presence of neutrino decay, the disappearance probability decreases. Around the oscillation peaks, the difference is significant, especially in T2HKK. This is clear from the following two-generation survival probability in a vacuum \cite{Chakraborty:2020cfu}:

\begin{equation}
P_{\mu \mu}=[1-\sin^2\theta_{23}(1-e^{-\frac{m_3 L}{\tau_3 E}})]^2-\sin^2{2\theta_{23}}e^{-\frac{m_3 L}{\tau_3 2E}}\sin^2(\frac{\Delta m^2_{31} L}{4E})
\end{equation}

In the case when there is no decay, the probability is determined by ${\sin^2}2\theta_{23}$. However, when decay is present, the $e^{-\frac{m_3 L}{\tau_3 E}}$ factor becomes significant and causes octant sensitivity.

\subsection{$\chi^2$ analysis}
\subsubsection{Constraints on Invisible Neutrino Decay}

\noindent
\begin{figure}[htbp]
\begin{center}
\includegraphics[scale=0.52]{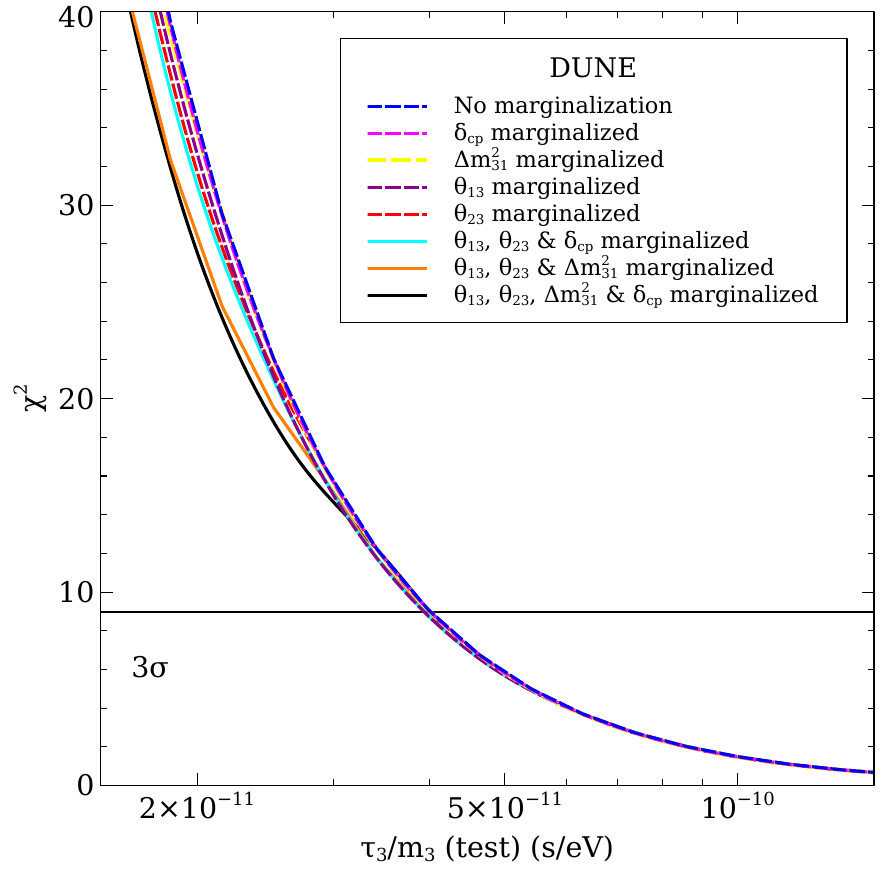}
\includegraphics[scale=0.52]{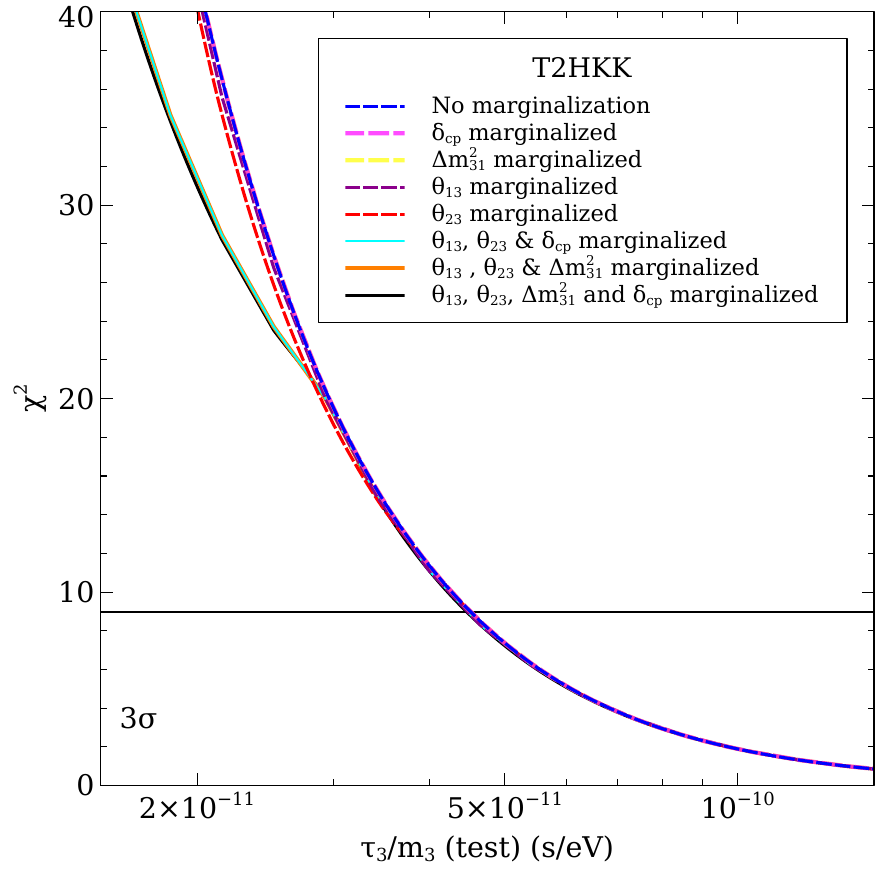}
\caption{The sensitivity $\chi^{2}$ as a function of ${\tau_3}/{m_3}$ for DUNE and T2HKK for different marginalization effects. The left panel is for DUNE and the right panel is for T2HKK. The data have been generated for the no-decay scenario.}
\label{fig: chi-marg}
\end{center}
\end{figure}

In this section, we perform a $\chi^2$ analysis to show how these experiments and their combinations can constrain invisible neutrino decay. We have  estimated the statistical $\chi^{2}$ function using the following relation:
 \begin{equation}
\chi^2_{stat}=2\sum_{i}\left[ N^{test}_{i}-N^{true}_{i}+N^{true}_{i}\log\frac{N^{true}_{i}}{N^{test}_{i}}\right]
\label{eq: chi}
\end{equation}
 Where $N^{true}_{i}$ represents the number of true events and $N^{test}_{i}$ represents the number of test events. True events are generated for the given true values in Table \ref{tab: 3sigma} and the $\chi^2$ is computed for a given set of true values by marginalizing over the test parameters in their $3\sigma$ allowed ranges. The uncertainties mentioned in Table \ref{tab: taberror} are incorporated in the $\chi^2$ by ``pull'' method using the following expression:

 \begin{equation}
 N^{\rm test}_i = \sum_{s(b)}  N^{s(b)}_i \bigg(1+ {c_i^{s(b)}}^{norm} {\xi^{s(b)}}^{norm} +  {c_i^{s(b)}}^{tilt} { \xi^{s(b)}}^{tilt} \frac{E_i - \bar{E}}{E_{max} - E_{min}}
\bigg)
 \end{equation} 
Here in this equation, $s(b)$ denotes signal(background), $\xi^{norm}$(${ \xi}^{tilt}$) is the ``pull'' variable and $c_i^{norm}$(${c_i}^{tilt}$) represents the change in the number of events due to the variation of the ``pull'' variable. In the above equation, $E_i$ is the mean reconstructed energy of the $i^{th}$ bin and the mean energy is given by, $\bar{E} = ({E_{max} +E_{min}})/{2}$ ($E_{min}$ and $E_{max}$ respectively are the maximum and minimum energy of the energy range).

Here we present the $\chi^2$ sensitivity as a function of the decay parameter. We explore the effect of different marginalization while constraining invisible neutrino decay.  The results are shown in Figure \ref{fig: chi-marg}. The data in all the cases shown in Figure \ref{fig: chi-marg} are generated with the standard oscillation parameters given in Table \ref{tab: 3sigma}. The true value of $\theta_{23}$ is considered in the higher octant, i.e., $\theta_{23} = 48.44^{\circ}$. The blue dashed lines represent the scenario where we have not marginalized any parameters. The dashed magenta, yellow, purple, and red lines represent the marginalization of  $\delta_{\mathrm{CP}}$, $\Delta m^2_{31}$, $\theta_{13}$, and $\theta_{23}$ respectively in their 3$\sigma$ allowed range. The solid cyan and brown lines represent the cases where the marginalization is done over $\theta_{13}$,  $\theta_{23}$, $\delta_{\mathrm{CP}}$ and $\theta_{13}$,  $\theta_{23}$, $\Delta{m}^2_{31}$  respectively. Finally, the black lines show the effect of marginalization over $\theta_{13}$, $\theta_{23}$, $\Delta{m}^2_{31}$ and $\delta_{\mathrm{CP}}$.

\begin{figure}[!h]
\begin{center}

\includegraphics[scale=0.54]{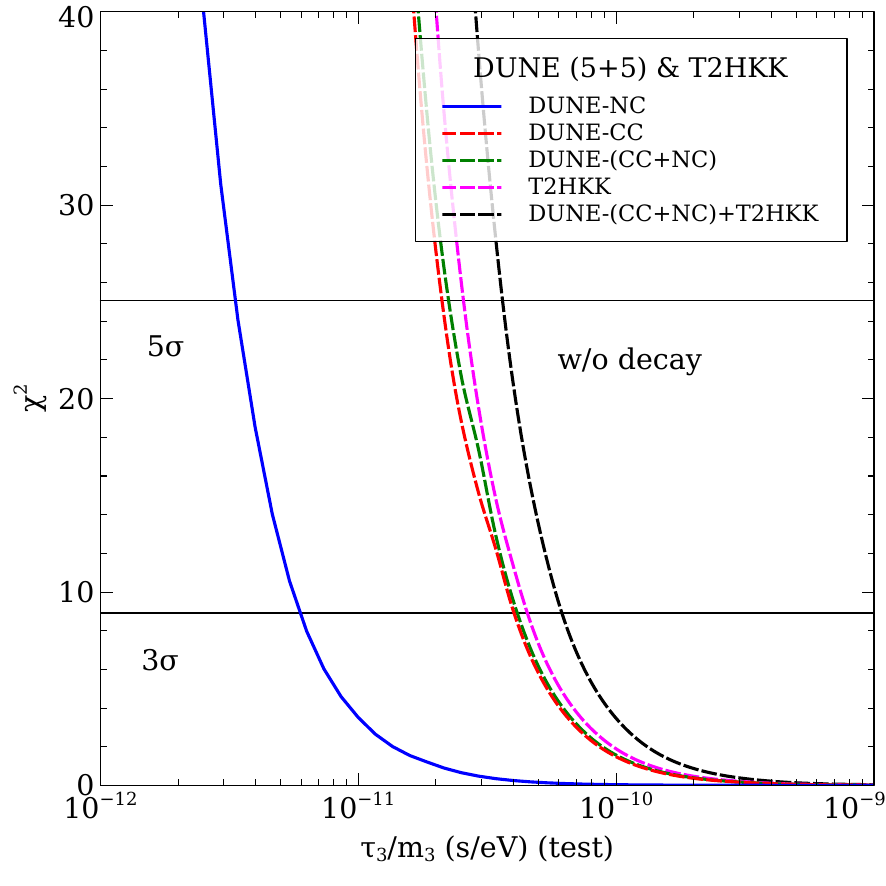}
\includegraphics[scale=0.54]{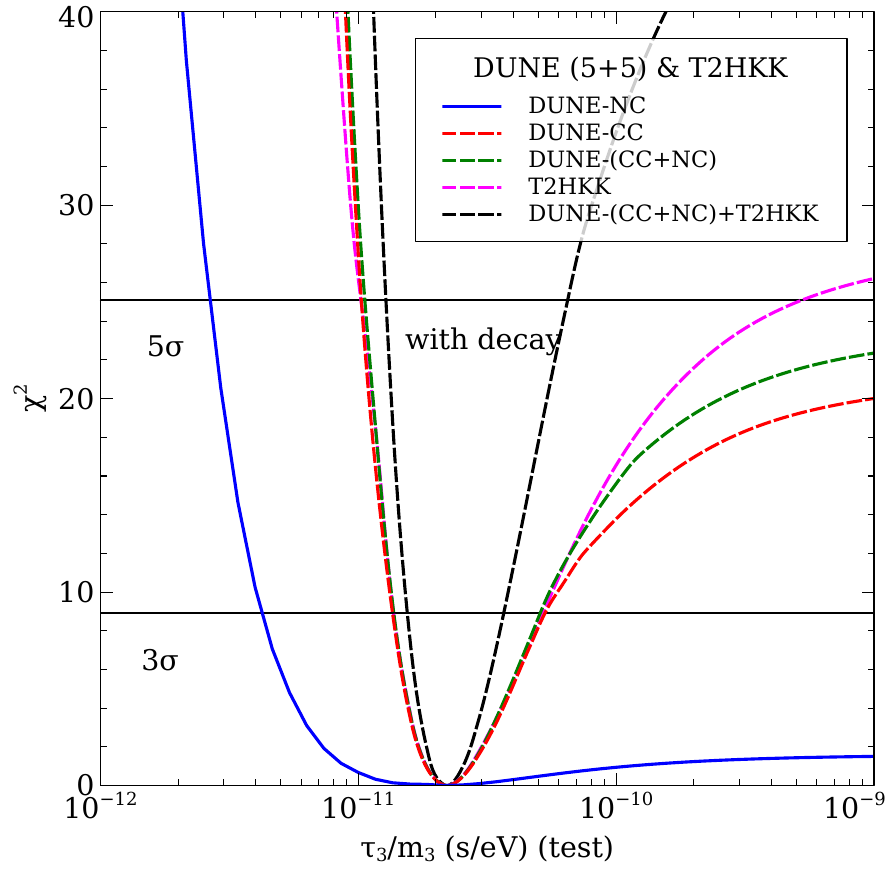}

\caption{$\chi^{2}$ as a function of ${\tau_3}/{m_3}$ for standalone and combination of experiments. The left panel is for no-decay in data and the right panel is for ${\tau_3}/{m_3}=2.2\times10^{-11}$ s/eV in data.}
\label{fig: chi-comb}
\end{center}
\end{figure}

In Figure \ref{fig: chi-marg}, we observe that at DUNE, the dashed blue line aligns closely with both the dashed magenta and yellow lines, indicating that marginalization over $\Delta m^2_{31}$ and $\delta_{\mathrm{CP}}$ has minimal impact on constraining invisible neutrino decay. The effects primarily arise from variations in $\theta_{13}$ and $\theta_{23}$. Again, marginalizing over $\theta_{13}$, $\theta_{23}$, and $\Delta m^2_{31}$ (represented by the brown line) significantly influences the constraint, with further adjustments possible upon inclusion of $\delta_{\mathrm{CP}}$. Similarly, for T2HKK, marginalizing over $\Delta{m}^2_{31}$ and $\delta_{\mathrm{CP}}$ shows negligible effects, while marginalization over $\theta_{13}$  influences the constraint. However, marginalization over $\theta_{23}$ shows observable effects. Notably, when considering constraints obtained by marginalizing over different combinations of parameters, such as $\theta_{13}$, $\theta_{23}$, $\Delta m^2_{31}$ (represented by the brown lines), $\theta_{13}$, $\theta_{23}$, and $\delta_{\mathrm{CP}}$ ( represented by the cyan lines), and $\theta_{13}$, $\theta_{23}$, $\Delta m^2_{31}$, and $\delta_{\mathrm{CP}}$ (represented by the black lines), they all exhibit almost identical behaviour. Hence, unless otherwise specified, we adopt marginalization over $\theta_{13}$, $\theta_{23}$, $\Delta m^2_{31}$, and $\delta_{\mathrm{CP}}$ for both experiments throughout the remainder of this study. \\

In Figure \ref{fig: chi-comb}, we show the sensitivities of DUNE, T2HKK, and their combinations to invisible neutrino decay. The results are shown for DUNE-NC, DUNE-CC, DUNE-(CC+NC), T2HKK, and DUNE-(CC+NC)+T2HKK. We consider the true value of $\theta_{23}$ in the higher octant (HO), which is 48.44$^\circ$. The blue solid lines in Figure \ref{fig: chi-comb} represent DUNE-NC, the red dashed lines represent DUNE-CC,  the magenta dashed lines represent T2HKK, the green dashed lines represent DUNE-(CC+NC), and the dashed black lines represent the combination of DUNE-(CC+NC) and T2HKK. In the left panel, the data is generated with the value of the oscillation parameters as given in Table \ref{tab: 3sigma}. These plots show how well DUNE, T2HKK, and their combinations can constrain invisible neutrino decay. In the left panel, we see that the inclusion of the neutral current with the charged current slightly improves the limits on the decay parameter. Additionally, the combination of DUNE-(CC+NC)+T2HKK provides the highest limit on the decay parameter. After 10 years (5 years of $\nu$ and 5 years of $\bar{\nu}$) of running of DUNE, the limits obtained by DUNE-CC, DUNE-(CC+NC), and DUNE-(CC+NC)+T2HKK are $4.07\times10^{-11}$ s/eV,  $4.22\times10^{-11}$ s/eV and $6.21\times10^{-11}$ s/eV, respectively, at a $3\sigma$ C.L. Notably, the constraint obtained from T2HKK is $4.39\times10^{-11}$ s/eV at the same C.L. which is better than DUNE-(CC+NC). \\

The right panel of Figure \ref{fig: chi-comb} shows how these experiments and their combinations can exclude the present standard oscillation scenario if invisible neutrino decay is a natural phenomenon with ${\tau_3}/m_3= 2.2\times10^{-11}$ s/eV. The true value of ${\tau_3}/m_3$ considered in this work is within the allowed region of T2K, NO$\nu$A and  MINOS/MINOS+  data analysis \cite{Ternes:2024qui}. So if $\nu_{3}$ is unstable in nature with a decay width ${\tau_3}/m_3= 2.2\times10^{-11}$ s/eV, then DUNE, after running for 10 years (5 years of $\nu$ and 5 years of $\bar{\nu}$), can exclude no-decay scenario at 3$\sigma$ C.L., whereas T2HKK alone can exclude the same at more than 5$\sigma$ C.L. The synergy between DUNE and T2HKK can further improve the exclusion limit to 5$\sigma$ C.L. It can be seen from the Figure \ref{fig: chi-comb} that, T2HKK has better precision than DUNE. For a given true value of ${\tau_3}/m_3=2.2\times10^{-11}$ s/eV, T2HKK can measure the decay parameter within the range $[5.31\times10^{-11}> \tau_{3}/m_{3} > 1.35\times10^{-11}]$ s/eV at $3\sigma$ C.L. On the other hand, DUNE-(CC+NC) can precisely measure the decay parameter within the range $[5.16\times10^{-11}>\tau_{3}/m_{3}>1.35\times10^{-11}]$ s/eV at 3$\sigma$ C.L. for the same value of decay parameter. 
The synergy between DUNE and T2HKK improves the precision further and can measure the decay parameter within the range $[6.44\times10^{-11}>\tau_{3}/m_{3}>1.28\times10^{-11}]$ s/eV $([3.64\times10^{-11}>\tau_{3}/m_{3}>1.55\times10^{-11}]$ s/eV) at 5$\sigma$ (3$\sigma$) C.L.

\subsubsection{Measurements of $\theta_{23}$ in the presence of Invisible Neutrino Decay}

\begin{figure}[htbp]
\begin{center}
\includegraphics[scale=0.48]{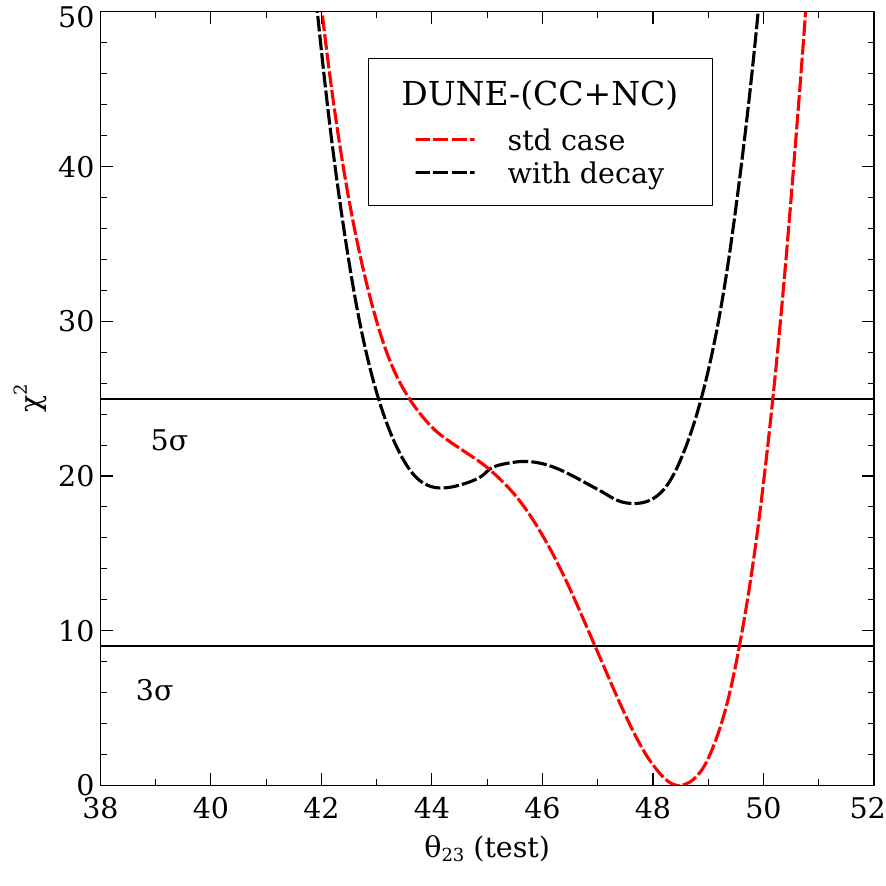}
\includegraphics[scale=0.48]{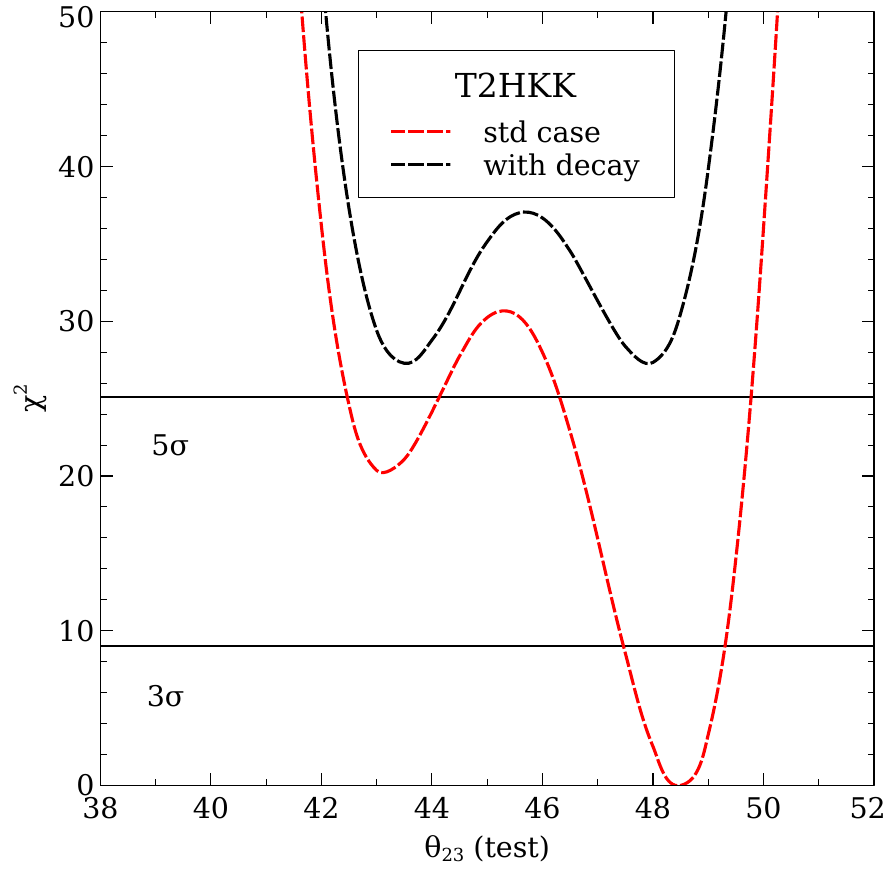}
 
\end{center}
\begin{center}

\includegraphics[scale=0.48]{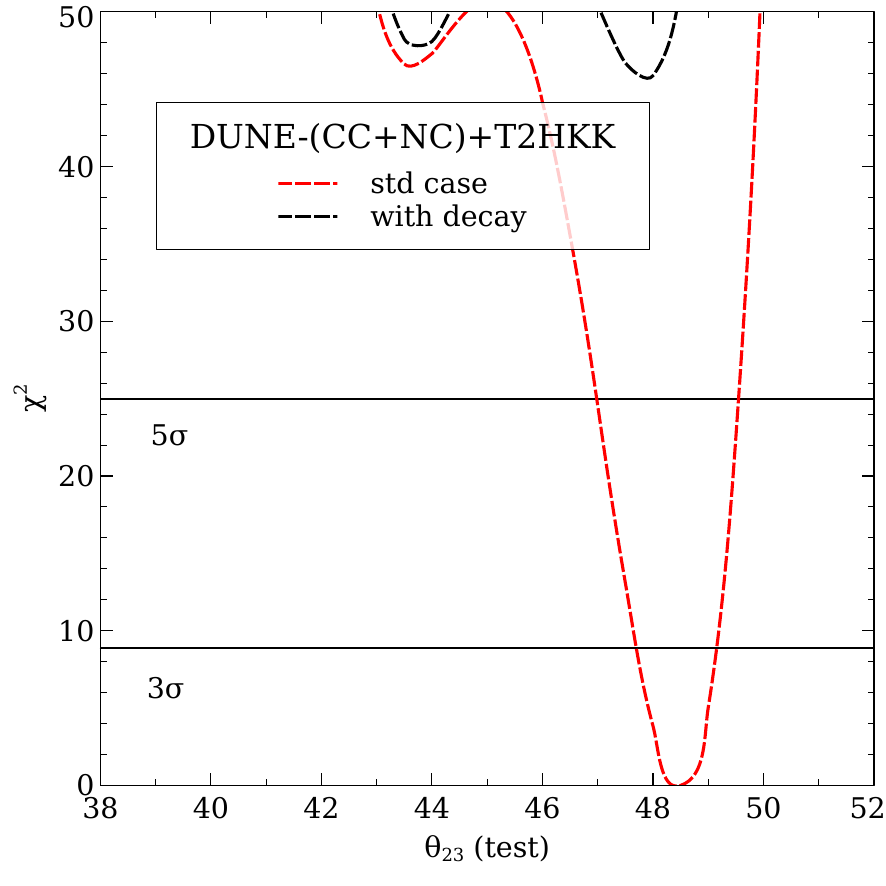} 
\end{center}

\caption{${\chi}^2$  as  a function of sin$^2\theta_{23}$ (test). The top left panel corresponds to DUNE-(CC+NC), the top right panel corresponds to the combination of both experiments. The dashed red line is for the stable neutrino case while the dashed black line represents the decay scenario. }
\label{fig: testtheta23}
\end{figure}

In this section, we show the effect of invisible neutrino decay in the measurements of $\theta_{23}$. We show the results considering $\theta_{23}=48.44^{\circ}$ at the higher octant. In the standard case, the true value of the decay parameter is zero while in the presence of invisible neutrino decay, the considered true value of the decay parameter is ${\tau_3}/{m_3}=2.2\times10^{-11}$ s/eV. Figure \ref{fig: testtheta23} shows the results of this analysis. The dashed red line represents the standard scenario while the dashed black line represents the decay scenario (non-zero decay in data). In the fit we have marginalized over $\theta_{13}$, $\Delta m^{2}_{31}$ and $\delta_{\mathrm{CP}}$ in their $3\sigma$ allowed range.

%\subsection{ ${\chi}^2$  as a function of sin$^2\theta_{23}$}

  From this analysis, we have observed that the position of the global minima changes in both the octants if invisible neutrino decay is a natural phenomenon. In the case of DUNE, (Figure \ref{fig: testtheta23}), the global true minima for the stable neutrino scenario is situated at the higher octant (HO) at $\theta_{23}= 48.44^{\circ}$ (as expected) while the fake minima appear in the LO at $\theta_{23} = 44.07^{\circ}$. However, the inclusion of neutrino decay alters the positions of these minima in both the octants. At DUNE, the $\chi^2$ minima shifts to $\theta_{23} = 47.72^\circ$ in the HO, while in the LO, fake minima appears at $\theta_{23} = 44.16^{\circ}$. This behaviour is consistent in both DUNE and T2HKK, as well as in their combined analyses. In T2HKK, in the presence of neutrino decay, the minima in the HO occur at $\theta_{23} = 47.89^{\circ}$, and fake minima appear in LO at $\theta_{23} = 43.62^{\circ}$. For the combination of DUNE and T2HKK, the minima in the HO shifts to $\theta_{23} = 47.86^{\circ}$ and in the LO, it appears at $\theta_{23} = 43.79^{\circ}$.
 This shift in the positions of $\chi^2$ minima is comprehensively discussed in \cite{Choubey:2017dyu, Chakraborty:2020cfu}.
 
\subsubsection{Octant of $\theta_{23}$ and CP violation Sensitivity in the presence of Invisible Neutrino Decay}

\noindent
\begin{figure}[htbp]
\begin{center}

\includegraphics[scale=0.48]{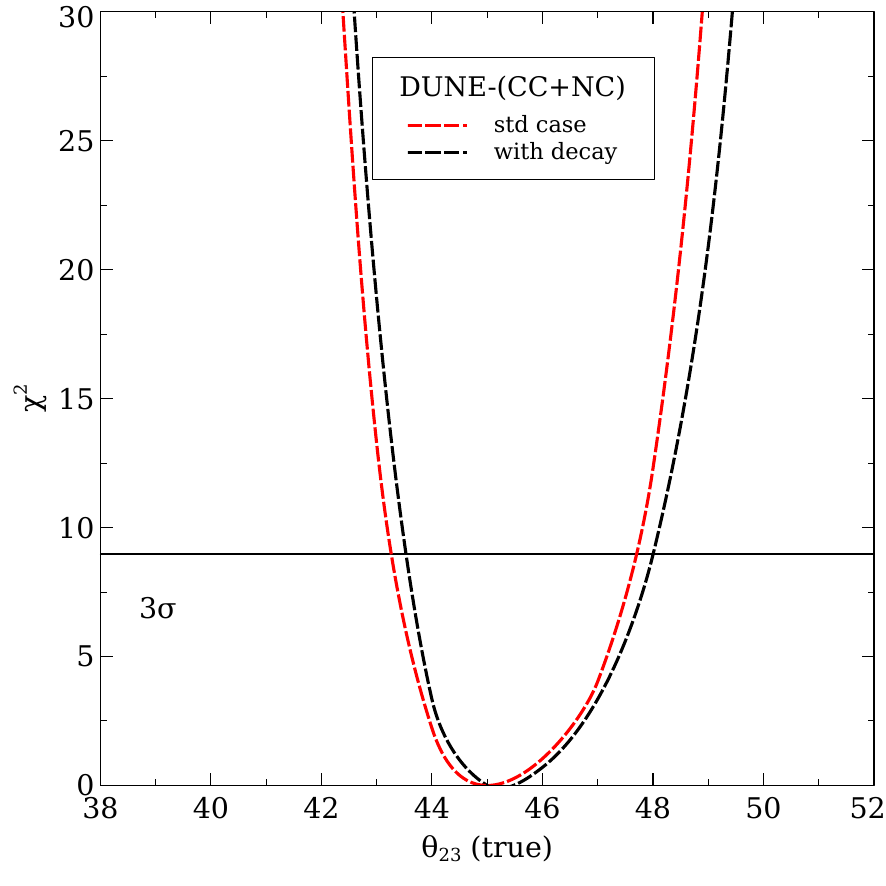}
\includegraphics[scale=0.48]{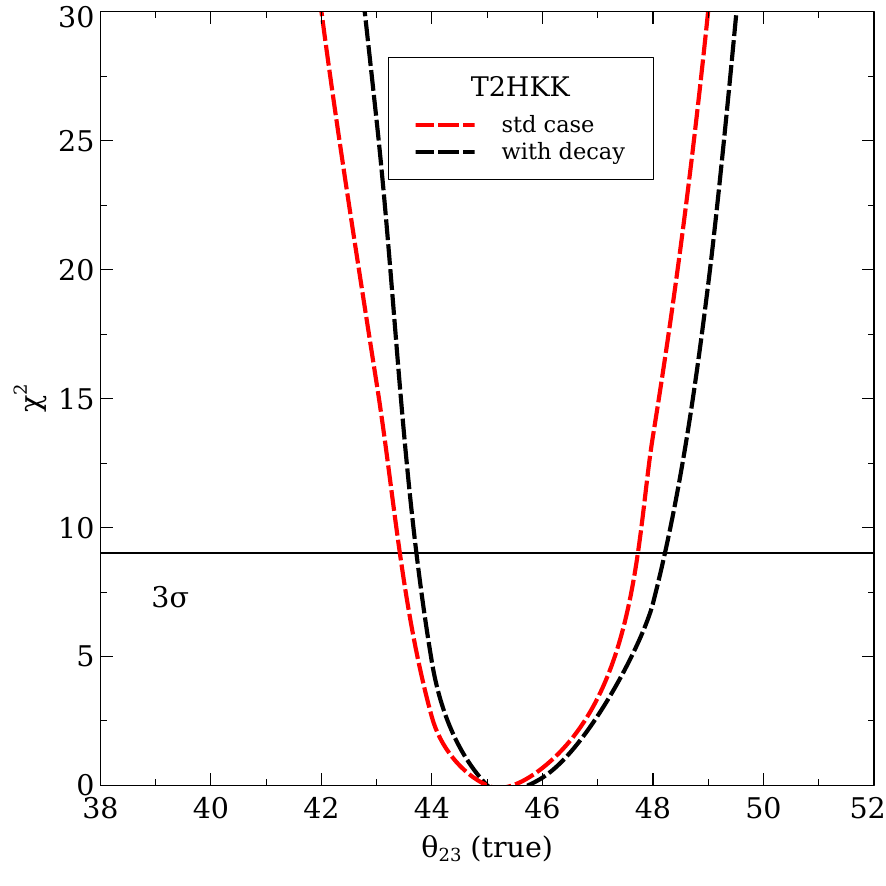}
\end{center}
\begin{center}

\includegraphics[scale=0.48]{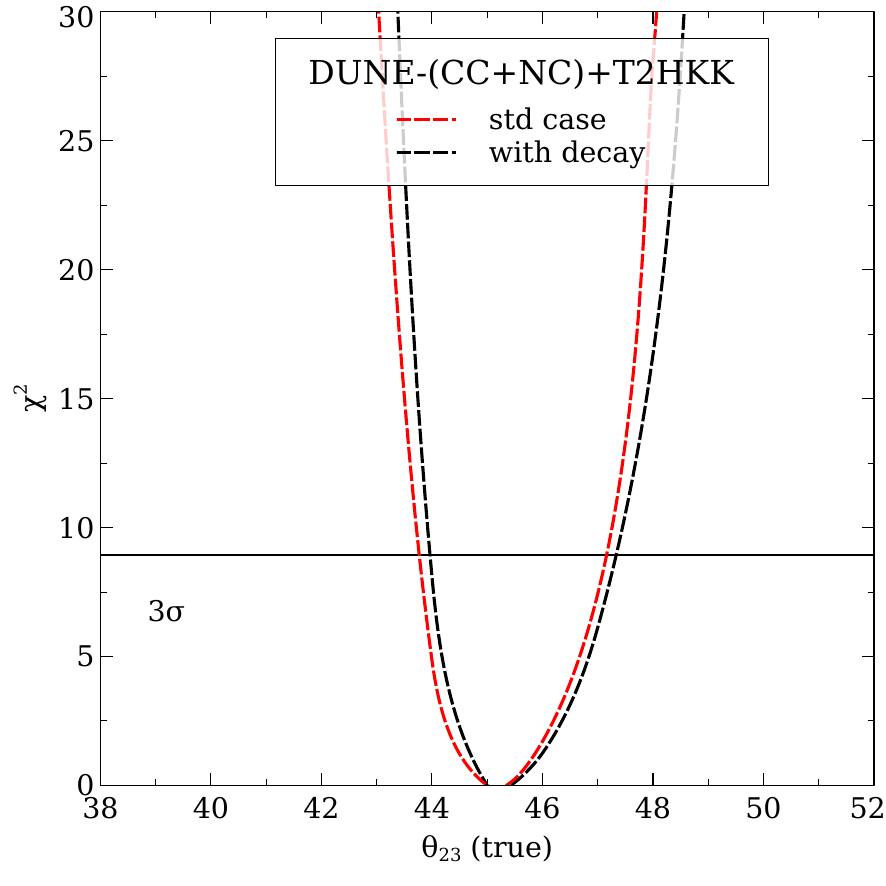} 
\end{center} 

\caption{Expected octant-sensitivity at DUNE-(CC+NC) top left, T2HKK top right , and DUNE-(CC+NC)+T2HKK bottom panel. The red dashed lines are for standard oscillation case and the black dashed lines are in the presence of invisible neutrino decay.}
\label{fig: octant}
\end{figure}

\begin{figure}[htbp]
\begin{center}
\includegraphics[scale=0.55]{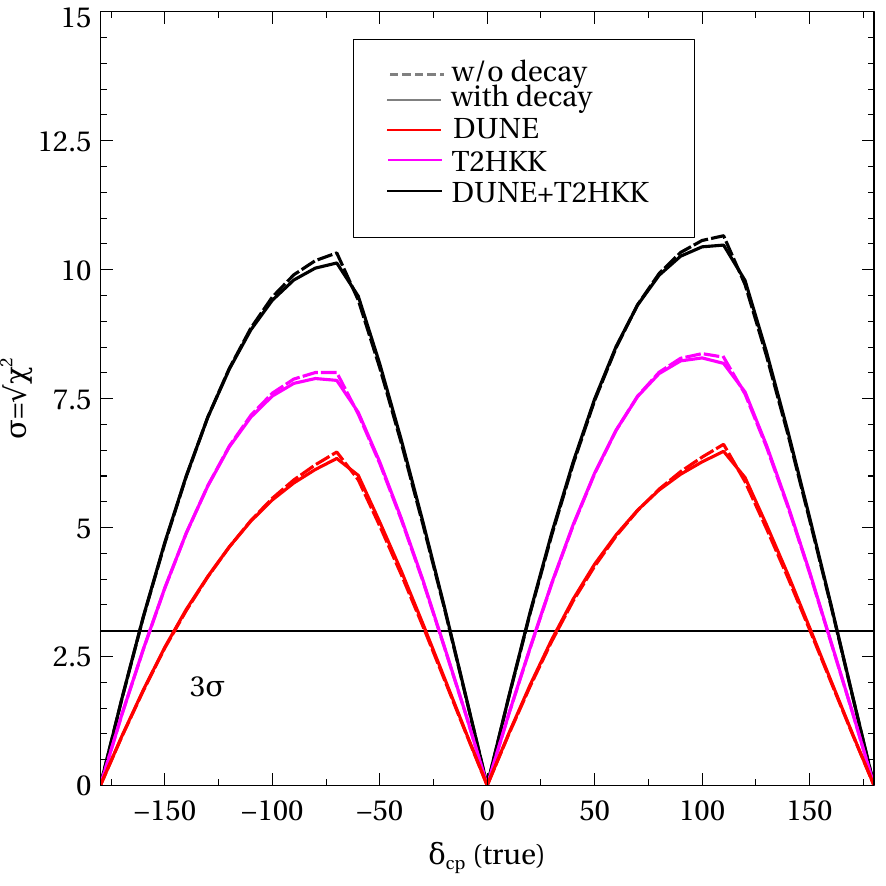} 

\caption{CP violation sensitivity for  DUNE-(CC+NC) (red), T2HKK (magenta), and DUNE-(CC+NC)+T2HKK (black) experiments. The dashed lines are for the standard oscillation case and the solid lines are for the decay case.}
\label{fig: cp}
\end{center}
\end{figure}

In this section, we study the effect of invisible neutrino decay in the measurements of octant and CP violation sensitivity at the far detector of these experiments.
Figure \ref{fig: octant} and \ref{fig: cp} show the octant resolution capacity and the CP violation sensitivity for DUNE-(CC+NC), T2HKK, and the combination of DUNE-(CC+NC) and T2HKK respectively.
To address the ability of the considered experiments to rule out the wrong octant, we vary the true $\theta_{23}$ in the range [$38^{\circ}, 52^{\circ}$], keeping the true values of all other oscillation parameters fixed at their best-fit values( as given in \ref{tab: 3sigma}). In the presence of invisible neutrino decay, a non-zero value of the decay parameter ($\tau_3/m_3=2.2\times10^{-11}$ s/eV) is considered in the data. We show the $\chi^2$ as a function of true $\theta_{23}$ assuming NH as the true hierarchy. In the fit we marginalize over $\theta_{13}, \Delta{m}^2_{31}$ and $\delta_{\mathrm{CP}}$ in their allowed $3\sigma$ range. Also, for every true value of $\theta_{23}$ in the lower octant (LO), we vary the test $\theta_{23}$ in the higher octant (HO) and vice-versa. In Figure \ref{fig: octant}, the dashed red line represents the standard oscillation scenario while the dashed black line represents the decay scenario. If neutrino decay is a natural occurrence, it will change the capabilities of these experiments to resolve octant degeneracy. In standard oscillation case, DUNE (T2HKK) can fix octant degeneracy for all true $\theta_{23} \le 43.28^{\circ}$ ($\theta_{23} \le 43.39^{\circ}$) and $\theta_{23}\ge 47.72^{\circ}$ ($\theta_{23}\ge 47.75^{\circ}$) at $3\sigma$. In this case, the combination of DUNE(CC+NC) and T2HKK can rule out the wrong octant at $3\sigma$ for all true $\theta_{23} \le 43.76^{\circ}$ and $\theta_{23}\ge 47.18^{\circ}$. At DUNE (T2HKK), in the presence of neutrino decay, octant resolution is possible for all true $\theta_{23} \le 43.53^{\circ}$ ($\theta_{23} \le 43.73^{\circ}$) and $\theta_{23}\ge 48^{\circ}$ ($\theta_{23}\ge 48.23^{\circ}$) at $3\sigma$. The combination can resolve octant at $3\sigma$ for all true  $\theta_{23} \le 43.99^{\circ}$ and $\theta_{23}\ge 47.38^{\circ}$ respectively. Therefore, in comparison to the standard oscillation scenario, the octant resolving capabilities of these setups slightly improve in the LO but deteriorate in the HO in the presence of invisible neutrino decay. Based on the foregoing analysis, we find that in the presence of invisible neutrino decay, the combined effect of DUNE and T2HKK can exclude the wrong octant somewhat more effectively than either experiment alone.

%In this section, we study the effect of neutrino decay in the measurements of octant and CP violation sensitivity at the far detector of these experiments.Figure \ref{fig: octant} and \ref{fig: cp} show the octant sensitivity and CP violation sensitivity for DUNE-(CC+NC), T2HKK, and the combination of DUNE-(CC+NC) and T2HKK respectively. In Figure \ref{fig: octant}, the dashed red lines represent the standard oscillation scenario while the dashed black lines represent the decay scenario. We assume the value of decay parameter $\tau_3/m_3=2.2\times10^{-11}$ s/eV. We generate the data at the values of oscillation parameters given in Table \ref{tab: 3sigma}. While analyzing the octant sensitivity in the lower octant (LO), for every true value of $\theta_{23}$ in the lower octant, we vary the test $\theta_{23}$ in the higher octant in the range [$45^{\circ}, 52^{\circ}$]. Similarly for any true value of $\theta_{23}$ in the higher octant (HO), the test $\theta_{23}$ is varied in LO in the range [$38^{\circ}, 45^{\circ}$]. In the fit we marginalize over $\theta_{13}, \Delta{m}^2_{31}$ and $\delta_{\mathrm{CP}}$ in their allowed $3\sigma$ range given in Table \ref{tab: 3sigma}. We observe that for all the cases in the presence of invisible neutrino decay, the octant sensitivity increases in the lower octant and decreases in the higher octant. %Furthermore, when we combine DUNE-(CC+NC) with T2HKK, the overall octant sensitivity increases.%

\noindent

In Figure \ref{fig: cp}, we present the CP violation sensitivity plots for DUNE-(CC+NC), T2HKK, and the combination of DUNE-(CC+NC) and T2HKK. The true value of $\theta_{23}$ considered in this case is 48.44$^{\circ}$. Here, all the solid lines represent the decay scenario, while the dashed lines represent the stable neutrino case. In Figure \ref{fig: cp} the red lines depict DUNE-(CC+NC), magenta lines represent  T2HKK, and the black lines represent DUNE-(CC+NC)+T2HKK.
 CP violation sensitivity of an experiment is the measure of its ability to exclude the CP-conserving values. To measure the CP violation sensitivity, we vary the true $\delta_{\mathrm{CP}}$ within the range $[-\pi, \pi]$ in data while in the fit, we fix $\delta_{\mathrm{CP}}$ at $0$ and $\pm\pi$. The oscillation parameters $\theta_{13}, \theta_{23}$ and $\Delta{m}^2_{31}$ are marginalized in their $3\sigma$ range keeping the other parameter fixed as given in Table \ref{tab: 3sigma}. We see that the invisible neutrino decay brings no noticeable change in the CP sensitivity plots.

\subsubsection{Correlation plots}

\noindent
\begin{figure}[htbp]
\begin{center}
\includegraphics[scale=0.54]{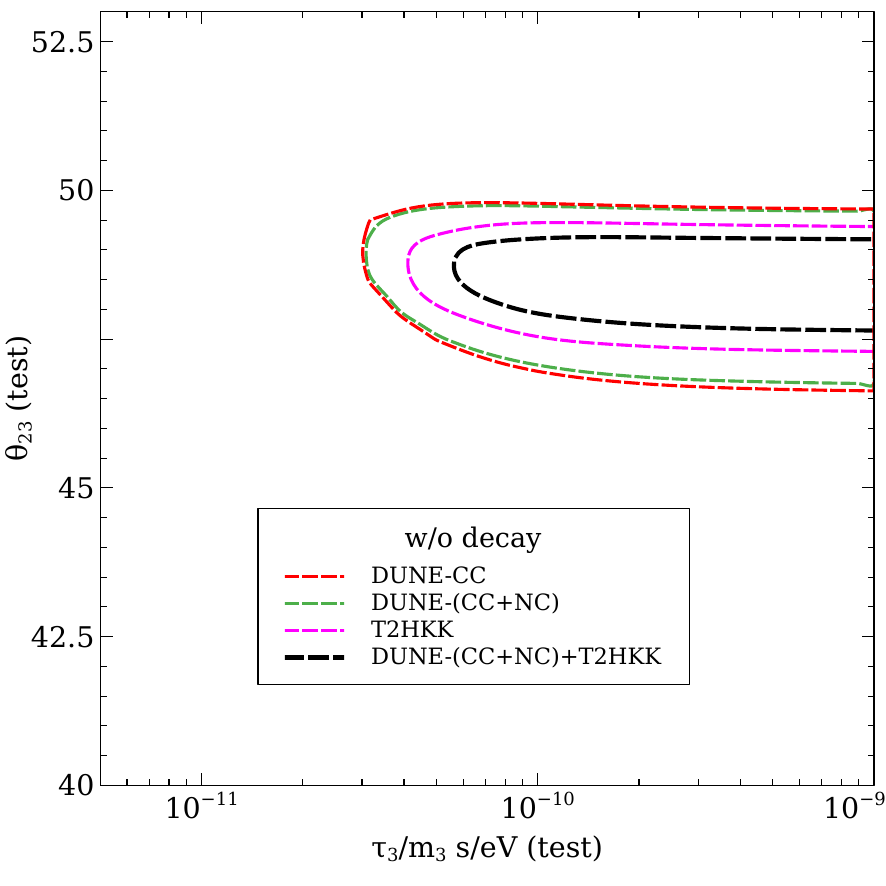} 
\includegraphics[scale=0.54]{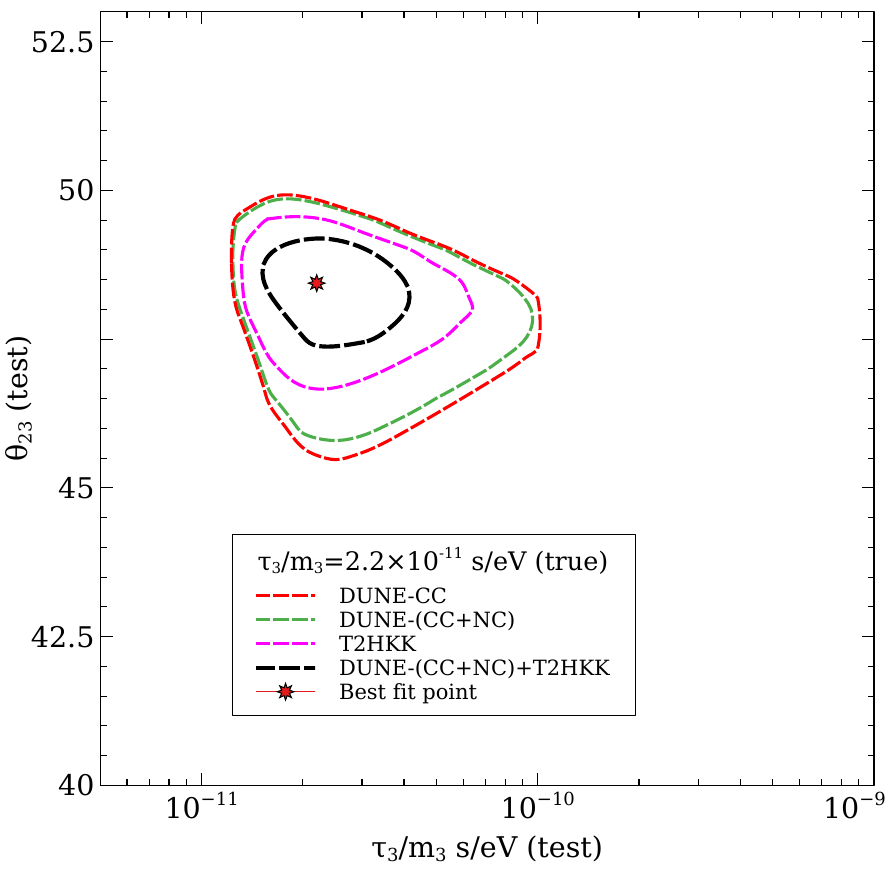} 
\caption{The $3\sigma$ confidence contours in the ${\tau_3}/{m_3}-\theta_{23}$ plane for DUNE (CC-red lines, CC+NC-green lines), T2HKK ( magenta lines) and DUNE-(CC+NC)+T2HKK (black lines). The left panel is for no-decay scenario and the right panel is for ${\tau_3}/{m_3}=2.2\times10^{-11}$ s/eV in data.}
\label{fig: alpha}
\end{center}
\end{figure}

In Figure \ref{fig: alpha}, we present the correlation between ${\tau_3}/{m_3}$ and $\theta_{23}$ at $3\sigma$ C.L. in ${\tau_3}/{m_3}-\theta_{23}$ plane. The results are generated for DUNE, T2HKK and for their combinations. In the left panel, we consider the no-decay scenario and in the right panel, we assume that neutrino decay is a natural phenomenon in addition to oscillations. In the analysis, we use the true value of the atmospheric mixing angle $\theta_{23}=48.44^{\circ}$ and the true value of the decay parameter  ${\tau_3}/{m_3}=2.2\times10^{-11}$ s/eV. This plot complements the results shown in Figure \ref{fig: chi-comb} but in ${\tau_3}/{m_3}-\theta_{23}$ plane. The smaller the parameter space, the more precise the measurements are. For DUNE, parameter space shrinks slightly in both panels when we add neutral current measurements to the charged current measurements. However, T2HKK performs better than DUNE in both scenarios. If neutrino decays in addition to oscillation, then T2HKK can significantly constrain the parameter space and hence can make more precise measurements in the 3$\sigma$ allowed region. The combinations of DUNE and T2HKK enhance the precision. In this case, the synergy between DUNE and T2HKK aids in all aspects to probe invisible neutrino decay.

%\subsection{The $3\sigma$ confidence contours in the sin$^2\theta_{23}-
%\Delta{m}^2_{31}$ plane}

\noindent
\begin{figure}[htbp]
\begin{center}
\includegraphics[scale=0.40]{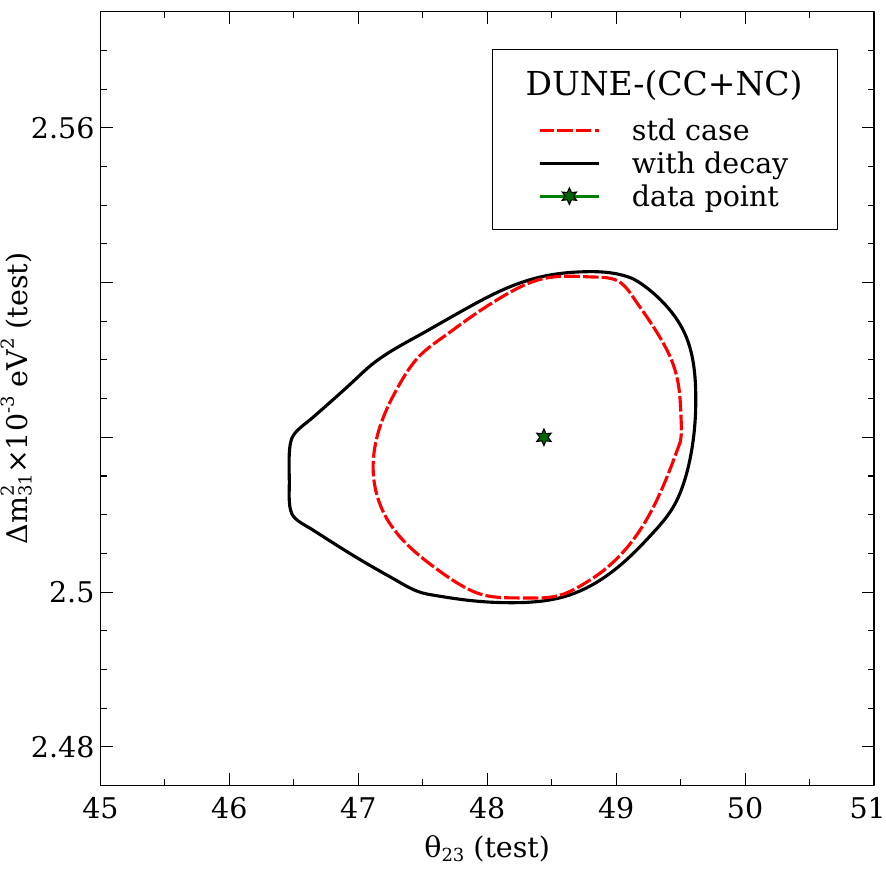}
\includegraphics[scale=0.40]{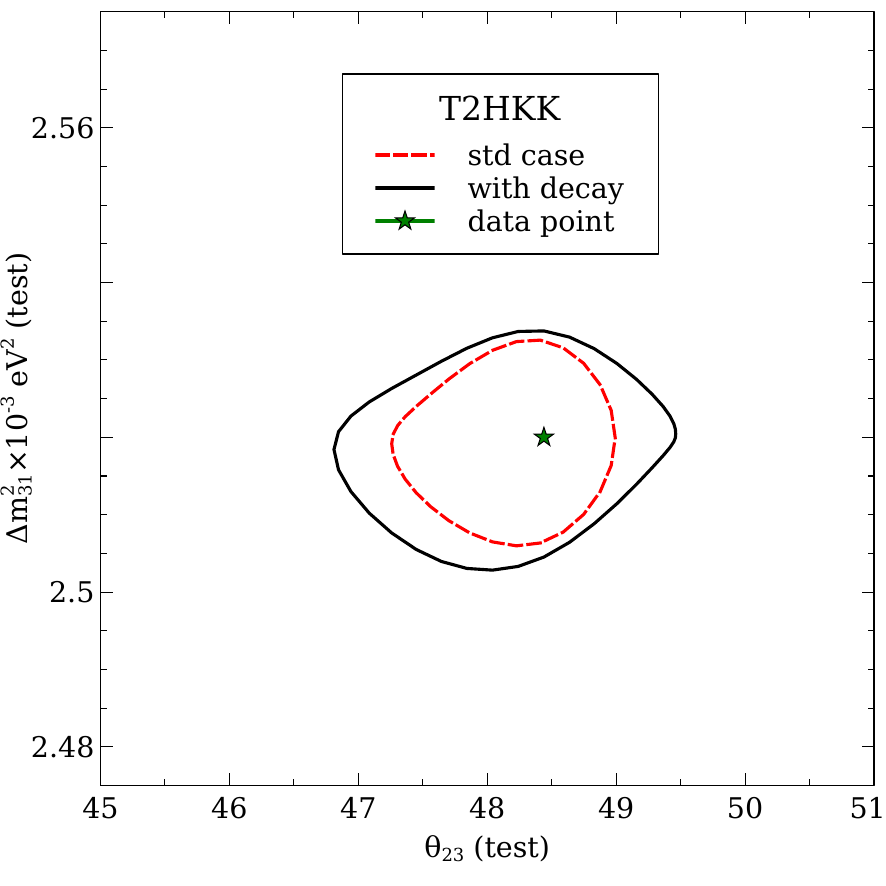}
\end{center}

\begin{center}

\includegraphics[scale=0.40]{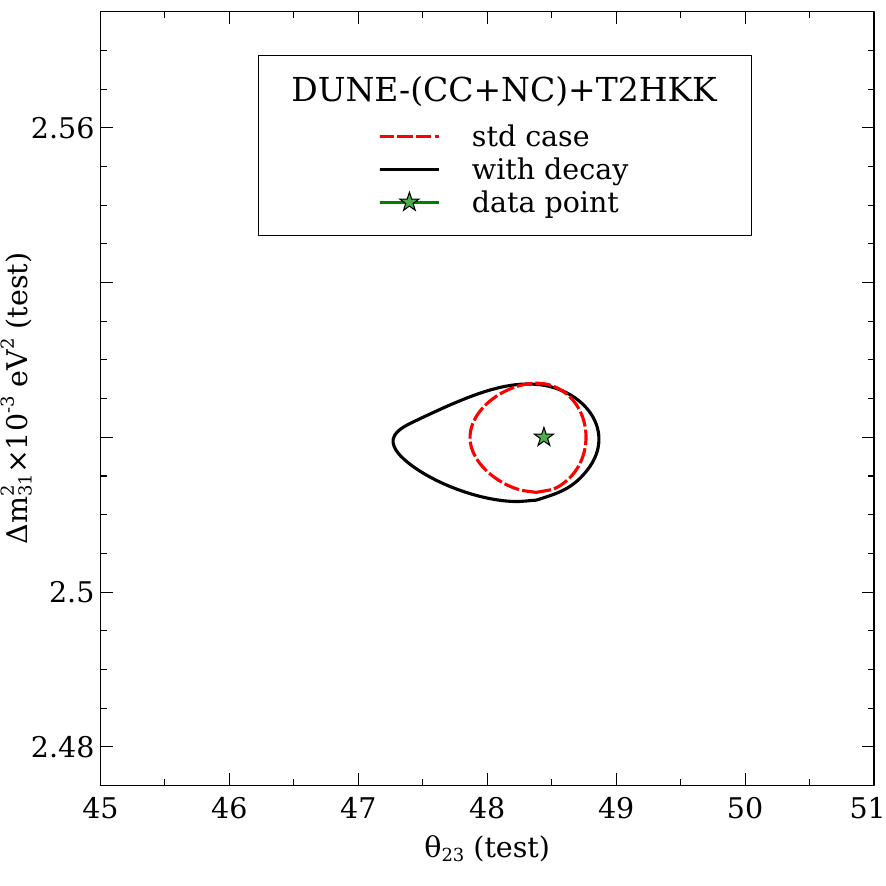} 
\end{center} 

\caption{The $3\sigma$ confidence contours in the $\theta_{23}-\Delta{m}^2_{31}$ plane for DUNE-(CC+NC) (top left), T2HKK  (top right), and DUNE-(CC+NC)+T2HKK (bottom) experiments. The red dashed contours are for standard oscillation case and solid black contours are for ${\tau_3}/{m_3}=2.2\times10^{-11}$ s/eV.}
\label{fig: dm}

\end{figure}

In Figure \ref{fig: chi-marg}, we have already observed the effects of $\theta_{23}$ and $\Delta{m}^2_{31}$ on constraining invisible neutrino decay. Both parameters play an important role in the presence of invisible neutrino decay. Here in Figure \ref {fig: dm}, we show the $3\sigma$ allowed region in $\theta_{23}-\Delta{m}^2_{31}$ plane for DUNE, T2HKK and for the combination of both the experiments, with and without neutrino decay. For all the cases in Figure \ref{fig: dm} the red dashed contours represent the stable neutrino case while the solid black contours represent the decay case. For the standard oscillation case, the contours are marginalized over $\theta_{13}$ and $\delta_{\mathrm{CP}}$ in their current $3\sigma$ ranges. While for the decay case, we have varied the decay parameter in the range $10^{-12}$ to $10^{-9}$ (s/eV) along with the marginalization of $\theta_{13}$ and $\delta_{\mathrm{CP}}$ in their current $3\sigma$ ranges. We observe that in the presence of invisible neutrino decay with oscillation, the allowed region increases more than the standard case, in all three cases. The precision of measurement at T2HKK is higher than that of DUNE as the $\theta_{23}-\Delta{m}^2_{31}$ parameter space shrinks in the case of T2HKK. The precision of measurements increases when we combine both experiments. In table \ref{tab: tabconstrain}, we present a comparison of the limit on decay parameters obtained from various experiments.

\newcolumntype{P}[1]{>{\centering\arraybackslash}p{2.5cm}}
\newcolumntype{P}[1]{>{\centering\arraybackslash}p{5.4cm}}
\begin{table}[htbp]
\renewcommand{\arraystretch}{2}
\begin{center}

\begin{tabular}{|P{5cm}|P{5cm}|P{5cm}|}

\hline
Experiment & $90\% $ C.L. $(3\sigma)$ bound on $\tau_3/m_3$ (s/eV)  & Ref. \\
\hline
SK+MINOS &  $2.9(0.54)\times10^{-10}$& Ref. \citep{Gonzalez-Garcia:2008mgl}\\
%\hline
T2K + NO$\nu$A & $2.3(1.5)\times10^{-12}$ & Ref. \cite{Choubey:2018cfz}\\
%\hline
T2K+NO$\nu$A+MINOS&$2.4(2.4)\times10^{-11}$& Ref. \citep{Ternes:2024qui}\\
\hline

DUNE - CC & $4.50 (2.38)\times 10^{-11}$&  Ref. \citep{Choubey:2017dyu} \\
%\hline
DUNE - (CC+NC) (5+5)& $5.1 (2.7)\times10^{-11}$& Ref. \citep{Ghoshal:2020hyo}\\
%\hline
T2HKK & $8.41\times10^{-11}(4.39\times10^{-11})$& This work\\
%\hline
DUNE - (CC+NC) (5+5) & $7.74 (4.22)\times 10^{-11}$& This work\\
%\hline
DUNE - (CC+NC) (5+5)+T2HKK  & $1.12 \times10^{-10}(6.21\times10^{-11})$& This work\\
%\hline
 ESSnuSB (540 km) & $4.22 (1.68)\times10^{-11}$& Ref. \citep{Choubey:2020dhw}  \\
%\hline
 ESSnuB (360 km) & $4.95 (2.64)\times 10^{-11}$ & Ref. \citep{Choubey:2020dhw} \\
%\hline
 JUNO& $9.3(4.7)\times 10^{-11}$& Ref. \citep{Abrahao:2015rba}\\
%\hline
INO& $1.51(0.566)\times 10^{-10}$& Ref. \citep{Choubey:2017eyg} \\
%\hline
KM3NeT-ORCA & $2.5(1.4)\times10^{-10}$ &Ref. \citep{deSalas:2018kri} \\
%\hline
T2HK & $4.43 (2.72)\times10^{-11}$& Ref. \citep{Chakraborty:2020cfu}\\
%\hline
T2HKK & $1.01\times10^{-10}(4.36\times10^{-11})$& Ref. \cite{Chakraborty:2020cfu}\\
%\hline
T2HKK+ESSnuSB & $ 1.064\times10^{-10}(5.53\times10^{-11})$& Ref. \citep{Chakraborty:2020cfu}\\
%\hline
ESSnuSB & $3.69(2.43)\times10^{-11}$& Ref. \citep{Chakraborty:2020cfu}\\
%\hline
T2HK+ESSnuSB&$1.01\times10^{-10}(4.36\times10^{-11})$& Ref. \citep{Chakraborty:2020cfu}\\
\hline
\end{tabular}
\caption{Comparision of bound on decay parameter ${\tau_3}/{m_3}$ from various experiments. The bounds for SK+MINOS, T2K+NO$\nu$A and T2K+NO$\nu$A+MINOS are based on actual observational data.}
\label{tab: tabconstrain}
\end{center}
\end{table}

\section{Summary and Conclusion}
\label{sec: summ}

Invisible neutrino decay could have a major role in cosmology. Assuming that the neutrinos are stable, the bound on the sum of neutrino masses is measured in the cosmological timescale which are much stronger than the bounds from oscillation experiments. If invisible neutrino decay is a natural phenomenon, it may alter the equation of the state of the Universe and, consequently, the CMB spectra. As a result, the bound on the sum of neutrino masses may become less stringent \cite{Chacko:2019nej, Escudero:2020ped}, but the allowed values of $\tau/m$ are much larger. Hence, if one observed an anomalous signal in oscillation experiments, it would have an important impact on Cosmology.

In this work, we have explored the impact of invisible neutrino decay in two upcoming long baseline experiments, DUNE and T2HKK. We assumed that the mass eigenstate $m_{3}$ is heavier than $m_{4}$ and hence $\nu_3$ being unstable can decay to a lighter sterile state. Throughout this analysis, we have assumed normal mass hierarchy as the true hierarchy. Here, we have combined the neutral current measurements and charged current measurements at DUNE to constrain invisible neutrino decay. We found that this combination of CC and NC enhances DUNE's ability to constrain ${\tau_3}/{m_3}$. We found that the synergy between DUNE and T2HKK can provide tight constraints on ${\tau_3}/{m_3}$.

Firstly, we have checked the effect of marginalization on constraining invisible neutrino decay.  It is observed that the marginalization of $\Delta{m}^2_{31}$ and $\delta_{\mathrm{CP}}$ have a minimal effect while that of $\theta_{13}$ and $\theta_{23}$ have a significant effect on invisible neutrino decay, especially at DUNE. Based on this analysis, we have considered the marginalization of $\theta_{13}$, $\theta_{23}$, $\Delta{m}^2_{31}$ and $\delta_{\mathrm{CP}}$ while deriving the constraint on invisible neutrino decay. We found that after taking data for 10 years, DUNE (with both CC and NC measurements) at 3$\sigma$ C.L. can constrain ${\tau_3}/{m_3}$ at $4.22\times10^{-11}$ s/eV. We have observed that the T2HKK experiment has a better ability to constrain invisible neutrino decay than DUNE. This constraint on invisible decay improves further when combined with the DUNE experiment and at 3$\sigma$ C.L., the bound obtained is $6.21\times10^{-11}$ s/eV. If nature has invisible neutrino decay in addition to oscillations and the true value of ${\tau_3}/m_3= 2.2\times10^{-11}$ s/eV, the synergy between these two experiments can rule out a stable neutrino scenario up to 5$\sigma$ C.L. In that case, the combination of DUNE and T2HKK can measure the decay parameter within the range $[6.44\times10^{-11}>\tau_{3}/m_{3}>1.28\times10^{-11}]$ s/eV. 

The impact of decay in $\theta_{23}$ measurement has also been studied. We have found that the $\theta_{23}$ measurements are hampered in the presence of invisible neutrino decay. If the data is generated with neutrino decay, both the true and fake minima deviate from the minima corresponding to the stable neutrino. If neutrino decay is a natural process, then the measurements of $\theta_{23}$ get affected which is also seen from the correlation plots in the ${\tau_3}/{m_3}-\theta_{23}$ plane. 

This study also reveals that in the presence of invisible neutrino decay, the ability of these experiments to rule out the wrong octant changes as compared to the stable neutrino scenario. The octant resolving capabilities of these setups slightly improve in the LO but deteriorate in the HO in the presence of invisible neutrino decay.
 The effect of invisible neutrino decay in the measurement of CP violation sensitivity is nominal.  

While the NC measurements at DUNE have no significant effect on invisible neutrino decay, combining it with the CC measurements provides higher precision. Furthermore, combining DUNE-(CC+NC) with T2HKK improves the constraints on the decay parameter.

\pagebreak
\def\bibsection{}  
{\textbf{\large References } }

\bibliography{References}

%merlin.mbs apsrev4-1.bst 2010-07-25 4.21a (PWD, AO, DPC) hacked
%Control: key (0)
%Control: author (72) initials jnrlst
%Control: editor formatted (1) identically to author
%Control: production of article title (-1) disabled
%Control: page (0) single
%Control: year (1) truncated
%Control: production of eprint (0) enabled
\begin{thebibliography}{56}%
\makeatletter
\providecommand \@ifxundefined [1]{%
 \@ifx{#1\undefined}
}%
\providecommand \@ifnum [1]{%
 \ifnum #1\expandafter \@firstoftwo
 \else \expandafter \@secondoftwo
 \fi
}%
\providecommand \@ifx [1]{%
 \ifx #1\expandafter \@firstoftwo
 \else \expandafter \@secondoftwo
 \fi
}%
\providecommand \natexlab [1]{#1}%
\providecommand \enquote  [1]{``#1''}%
\providecommand \bibnamefont  [1]{#1}%
\providecommand \bibfnamefont [1]{#1}%
\providecommand \citenamefont [1]{#1}%
\providecommand \href@noop [0]{\@secondoftwo}%
\providecommand \href [0]{\begingroup \@sanitize@url \@href}%
\providecommand \@href[1]{\@@startlink{#1}\@@href}%
\providecommand \@@href[1]{\endgroup#1\@@endlink}%
\providecommand \@sanitize@url [0]{\catcode `\\12\catcode `\$12\catcode
  `\&12\catcode `\#12\catcode `\^12\catcode `\_12\catcode `\%12\relax}%
\providecommand \@@startlink[1]{}%
\providecommand \@@endlink[0]{}%
\providecommand \url  [0]{\begingroup\@sanitize@url \@url }%
\providecommand \@url [1]{\endgroup\@href {#1}{\urlprefix }}%
\providecommand \urlprefix  [0]{URL }%
\providecommand \Eprint [0]{\href }%
\providecommand \doibase [0]{http://dx.doi.org/}%
\providecommand \selectlanguage [0]{\@gobble}%
\providecommand \bibinfo  [0]{\@secondoftwo}%
\providecommand \bibfield  [0]{\@secondoftwo}%
\providecommand \translation [1]{[#1]}%
\providecommand \BibitemOpen [0]{}%
\providecommand \bibitemStop [0]{}%
\providecommand \bibitemNoStop [0]{.\EOS\space}%
\providecommand \EOS [0]{\spacefactor3000\relax}%
\providecommand \BibitemShut  [1]{\csname bibitem#1\endcsname}%
\let\auto@bib@innerbib\@empty
%</preamble>
\bibitem [{\citenamefont {Abe}\ \emph {et~al.}(2017)\citenamefont {Abe} \emph
  {et~al.}}]{T2K:2017hed}%
  \BibitemOpen
  \bibfield  {author} {\bibinfo {author} {\bibfnamefont {K.}~\bibnamefont
  {Abe}} \emph {et~al.} (\bibinfo {collaboration} {T2K}),\ }\href {\doibase
  10.1103/PhysRevLett.118.151801} {\bibfield  {journal} {\bibinfo  {journal}
  {Phys. Rev. Lett.}\ }\textbf {\bibinfo {volume} {118}},\ \bibinfo {pages}
  {151801} (\bibinfo {year} {2017})},\ \Eprint
  {http://arxiv.org/abs/1701.00432} {arXiv:1701.00432 [hep-ex]} \BibitemShut
  {NoStop}%
\bibitem [{\citenamefont {Adamson}\ \emph {et~al.}(2016)\citenamefont {Adamson}
  \emph {et~al.}}]{NOvA:2016kwd}%
  \BibitemOpen
  \bibfield  {author} {\bibinfo {author} {\bibfnamefont {P.}~\bibnamefont
  {Adamson}} \emph {et~al.} (\bibinfo {collaboration} {NOvA}),\ }\href
  {\doibase 10.1103/PhysRevLett.116.151806} {\bibfield  {journal} {\bibinfo
  {journal} {Phys. Rev. Lett.}\ }\textbf {\bibinfo {volume} {116}},\ \bibinfo
  {pages} {151806} (\bibinfo {year} {2016})},\ \Eprint
  {http://arxiv.org/abs/1601.05022} {arXiv:1601.05022 [hep-ex]} \BibitemShut
  {NoStop}%
\bibitem [{\citenamefont {Abi}\ \emph {et~al.}(2020{\natexlab{a}})\citenamefont
  {Abi} \emph {et~al.}}]{DUNE:2020lwj}%
  \BibitemOpen
  \bibfield  {author} {\bibinfo {author} {\bibfnamefont {B.}~\bibnamefont
  {Abi}} \emph {et~al.} (\bibinfo {collaboration} {DUNE}),\ }\href {\doibase
  10.1088/1748-0221/15/08/T08008} {\bibfield  {journal} {\bibinfo  {journal}
  {JINST}\ }\textbf {\bibinfo {volume} {15}},\ \bibinfo {pages} {T08008}
  (\bibinfo {year} {2020}{\natexlab{a}})},\ \Eprint
  {http://arxiv.org/abs/2002.02967} {arXiv:2002.02967 [physics.ins-det]}
  \BibitemShut {NoStop}%
\bibitem [{\citenamefont {Abi}\ \emph {et~al.}(2020{\natexlab{b}})\citenamefont
  {Abi} \emph {et~al.}}]{DUNE:2020ypp}%
  \BibitemOpen
  \bibfield  {author} {\bibinfo {author} {\bibfnamefont {B.}~\bibnamefont
  {Abi}} \emph {et~al.} (\bibinfo {collaboration} {DUNE}),\ }\href@noop {} {\
  (\bibinfo {year} {2020}{\natexlab{b}})},\ \Eprint
  {http://arxiv.org/abs/2002.03005} {arXiv:2002.03005 [hep-ex]} \BibitemShut
  {NoStop}%
\bibitem [{\citenamefont {Abi}\ \emph {et~al.}(2020{\natexlab{c}})\citenamefont
  {Abi} \emph {et~al.}}]{DUNE:2020mra}%
  \BibitemOpen
  \bibfield  {author} {\bibinfo {author} {\bibfnamefont {B.}~\bibnamefont
  {Abi}} \emph {et~al.} (\bibinfo {collaboration} {DUNE}),\ }\href {\doibase
  10.1088/1748-0221/15/08/T08009} {\bibfield  {journal} {\bibinfo  {journal}
  {JINST}\ }\textbf {\bibinfo {volume} {15}},\ \bibinfo {pages} {T08009}
  (\bibinfo {year} {2020}{\natexlab{c}})},\ \Eprint
  {http://arxiv.org/abs/2002.03008} {arXiv:2002.03008 [physics.ins-det]}
  \BibitemShut {NoStop}%
\bibitem [{\citenamefont {Abi}\ \emph {et~al.}(2020{\natexlab{d}})\citenamefont
  {Abi} \emph {et~al.}}]{DUNE:2020txw}%
  \BibitemOpen
  \bibfield  {author} {\bibinfo {author} {\bibfnamefont {B.}~\bibnamefont
  {Abi}} \emph {et~al.} (\bibinfo {collaboration} {DUNE}),\ }\href {\doibase
  10.1088/1748-0221/15/08/T08010} {\bibfield  {journal} {\bibinfo  {journal}
  {JINST}\ }\textbf {\bibinfo {volume} {15}},\ \bibinfo {pages} {T08010}
  (\bibinfo {year} {2020}{\natexlab{d}})},\ \Eprint
  {http://arxiv.org/abs/2002.03010} {arXiv:2002.03010 [physics.ins-det]}
  \BibitemShut {NoStop}%
\bibitem [{\citenamefont {Abi}\ \emph {et~al.}(2021{\natexlab{a}})\citenamefont
  {Abi} \emph {et~al.}}]{DUNE:2020fgq}%
  \BibitemOpen
  \bibfield  {author} {\bibinfo {author} {\bibfnamefont {B.}~\bibnamefont
  {Abi}} \emph {et~al.} (\bibinfo {collaboration} {DUNE}),\ }\href {\doibase
  10.1140/epjc/s10052-021-09007-w} {\bibfield  {journal} {\bibinfo  {journal}
  {Eur. Phys. J. C}\ }\textbf {\bibinfo {volume} {81}},\ \bibinfo {pages} {322}
  (\bibinfo {year} {2021}{\natexlab{a}})},\ \Eprint
  {http://arxiv.org/abs/2008.12769} {arXiv:2008.12769 [hep-ex]} \BibitemShut
  {NoStop}%
\bibitem [{\citenamefont {Abed~Abud}\ \emph {et~al.}(2022)\citenamefont
  {Abed~Abud} \emph {et~al.}}]{DUNE:2022aul}%
  \BibitemOpen
  \bibfield  {author} {\bibinfo {author} {\bibfnamefont {A.}~\bibnamefont
  {Abed~Abud}} \emph {et~al.} (\bibinfo {collaboration} {DUNE}),\ }\href@noop
  {} {\  (\bibinfo {year} {2022})},\ \Eprint {http://arxiv.org/abs/2203.06100}
  {arXiv:2203.06100 [hep-ex]} \BibitemShut {NoStop}%
\bibitem [{\citenamefont {Abi}\ \emph {et~al.}(2020{\natexlab{e}})\citenamefont
  {Abi} \emph {et~al.}}]{DUNE:2020jqi}%
  \BibitemOpen
  \bibfield  {author} {\bibinfo {author} {\bibfnamefont {B.}~\bibnamefont
  {Abi}} \emph {et~al.} (\bibinfo {collaboration} {DUNE}),\ }\href {\doibase
  10.1140/epjc/s10052-020-08456-z} {\bibfield  {journal} {\bibinfo  {journal}
  {Eur. Phys. J. C}\ }\textbf {\bibinfo {volume} {80}},\ \bibinfo {pages} {978}
  (\bibinfo {year} {2020}{\natexlab{e}})},\ \Eprint
  {http://arxiv.org/abs/2006.16043} {arXiv:2006.16043 [hep-ex]} \BibitemShut
  {NoStop}%
\bibitem [{\citenamefont {Abe}\ \emph {et~al.}(2015)\citenamefont {Abe} \emph
  {et~al.}}]{Hyper-KamiokandeProto-:2015xww}%
  \BibitemOpen
  \bibfield  {author} {\bibinfo {author} {\bibfnamefont {K.}~\bibnamefont
  {Abe}} \emph {et~al.} (\bibinfo {collaboration} {Hyper-Kamiokande Proto-}),\
  }\href {\doibase 10.1093/ptep/ptv061} {\bibfield  {journal} {\bibinfo
  {journal} {PTEP}\ }\textbf {\bibinfo {volume} {2015}},\ \bibinfo {pages}
  {053C02} (\bibinfo {year} {2015})},\ \Eprint
  {http://arxiv.org/abs/1502.05199} {arXiv:1502.05199 [hep-ex]} \BibitemShut
  {NoStop}%
\bibitem [{\citenamefont {proto Collaboration}\ \emph
  {et~al.}(2018)\citenamefont {proto Collaboration}, \citenamefont {Abe},
  \citenamefont {Abe}, \citenamefont {Ahn}, \citenamefont {Aihara},
  \citenamefont {Aimi}, \citenamefont {Akutsu}, \citenamefont {Andreopoulos},
  \citenamefont {Anghel}, \citenamefont {Anthony} \emph
  {et~al.}}]{hyper2018physics}%
  \BibitemOpen
  \bibfield  {author} {\bibinfo {author} {\bibfnamefont {H.-K.}\ \bibnamefont
  {proto Collaboration}}, \bibinfo {author} {\bibfnamefont {K.}~\bibnamefont
  {Abe}}, \bibinfo {author} {\bibfnamefont {K.}~\bibnamefont {Abe}}, \bibinfo
  {author} {\bibfnamefont {S.}~\bibnamefont {Ahn}}, \bibinfo {author}
  {\bibfnamefont {H.}~\bibnamefont {Aihara}}, \bibinfo {author} {\bibfnamefont
  {A.}~\bibnamefont {Aimi}}, \bibinfo {author} {\bibfnamefont {R.}~\bibnamefont
  {Akutsu}}, \bibinfo {author} {\bibfnamefont {C.}~\bibnamefont
  {Andreopoulos}}, \bibinfo {author} {\bibfnamefont {I.}~\bibnamefont
  {Anghel}}, \bibinfo {author} {\bibfnamefont {L.}~\bibnamefont {Anthony}},
  \emph {et~al.},\ }\href@noop {} {\bibfield  {journal} {\bibinfo  {journal}
  {Progress of Theoretical and Experimental Physics}\ }\textbf {\bibinfo
  {volume} {2018}},\ \bibinfo {pages} {063C01} (\bibinfo {year}
  {2018})}\BibitemShut {NoStop}%
\bibitem [{\citenamefont {Baussan}\ \emph {et~al.}(2014)\citenamefont {Baussan}
  \emph {et~al.}}]{ESSnuSB:2013dql}%
  \BibitemOpen
  \bibfield  {author} {\bibinfo {author} {\bibfnamefont {E.}~\bibnamefont
  {Baussan}} \emph {et~al.} (\bibinfo {collaboration} {ESSnuSB}),\ }\href
  {\doibase 10.1016/j.nuclphysb.2014.05.016} {\bibfield  {journal} {\bibinfo
  {journal} {Nucl. Phys. B}\ }\textbf {\bibinfo {volume} {885}},\ \bibinfo
  {pages} {127} (\bibinfo {year} {2014})},\ \Eprint
  {http://arxiv.org/abs/1309.7022} {arXiv:1309.7022 [hep-ex]} \BibitemShut
  {NoStop}%
\bibitem [{\citenamefont {An}\ \emph {et~al.}(2016)\citenamefont {An} \emph
  {et~al.}}]{JUNO:2015zny}%
  \BibitemOpen
  \bibfield  {author} {\bibinfo {author} {\bibfnamefont {F.}~\bibnamefont {An}}
  \emph {et~al.} (\bibinfo {collaboration} {JUNO}),\ }\href {\doibase
  10.1088/0954-3899/43/3/030401} {\bibfield  {journal} {\bibinfo  {journal} {J.
  Phys. G}\ }\textbf {\bibinfo {volume} {43}},\ \bibinfo {pages} {030401}
  (\bibinfo {year} {2016})},\ \Eprint {http://arxiv.org/abs/1507.05613}
  {arXiv:1507.05613 [physics.ins-det]} \BibitemShut {NoStop}%
\bibitem [{\citenamefont {Ahmed}\ \emph {et~al.}(2017)\citenamefont {Ahmed}
  \emph {et~al.}}]{ICAL:2015stm}%
  \BibitemOpen
  \bibfield  {author} {\bibinfo {author} {\bibfnamefont {S.}~\bibnamefont
  {Ahmed}} \emph {et~al.} (\bibinfo {collaboration} {ICAL}),\ }\href {\doibase
  10.1007/s12043-017-1373-4} {\bibfield  {journal} {\bibinfo  {journal}
  {Pramana}\ }\textbf {\bibinfo {volume} {88}},\ \bibinfo {pages} {79}
  (\bibinfo {year} {2017})},\ \Eprint {http://arxiv.org/abs/1505.07380}
  {arXiv:1505.07380 [physics.ins-det]} \BibitemShut {NoStop}%
\bibitem [{\citenamefont {Aartsen}\ \emph {et~al.}(2014)\citenamefont {Aartsen}
  \emph {et~al.}}]{IceCube-PINGU:2014okk}%
  \BibitemOpen
  \bibfield  {author} {\bibinfo {author} {\bibfnamefont {M.~G.}\ \bibnamefont
  {Aartsen}} \emph {et~al.} (\bibinfo {collaboration} {IceCube-PINGU}),\
  }\href@noop {} {\  (\bibinfo {year} {2014})},\ \Eprint
  {http://arxiv.org/abs/1401.2046} {arXiv:1401.2046 [physics.ins-det]}
  \BibitemShut {NoStop}%
\bibitem [{\citenamefont {de~Salas}\ \emph {et~al.}(2019)\citenamefont
  {de~Salas}, \citenamefont {Pastor}, \citenamefont {Ternes}, \citenamefont
  {Thakore},\ and\ \citenamefont {T\'ortola}}]{deSalas:2018kri}%
  \BibitemOpen
  \bibfield  {author} {\bibinfo {author} {\bibfnamefont {P.~F.}\ \bibnamefont
  {de~Salas}}, \bibinfo {author} {\bibfnamefont {S.}~\bibnamefont {Pastor}},
  \bibinfo {author} {\bibfnamefont {C.~A.}\ \bibnamefont {Ternes}}, \bibinfo
  {author} {\bibfnamefont {T.}~\bibnamefont {Thakore}}, \ and\ \bibinfo
  {author} {\bibfnamefont {M.}~\bibnamefont {T\'ortola}},\ }\href {\doibase
  10.1016/j.physletb.2018.12.066} {\bibfield  {journal} {\bibinfo  {journal}
  {Phys. Lett. B}\ }\textbf {\bibinfo {volume} {789}},\ \bibinfo {pages} {472}
  (\bibinfo {year} {2019})},\ \Eprint {http://arxiv.org/abs/1810.10916}
  {arXiv:1810.10916 [hep-ph]} \BibitemShut {NoStop}%
\bibitem [{\citenamefont {Bahcall}\ \emph {et~al.}(1972)\citenamefont
  {Bahcall}, \citenamefont {Cabibbo},\ and\ \citenamefont
  {Yahil}}]{Bahcall:1972my}%
  \BibitemOpen
  \bibfield  {author} {\bibinfo {author} {\bibfnamefont {J.~N.}\ \bibnamefont
  {Bahcall}}, \bibinfo {author} {\bibfnamefont {N.}~\bibnamefont {Cabibbo}}, \
  and\ \bibinfo {author} {\bibfnamefont {A.}~\bibnamefont {Yahil}},\ }\href
  {\doibase 10.1103/PhysRevLett.28.316} {\bibfield  {journal} {\bibinfo
  {journal} {Phys. Rev. Lett.}\ }\textbf {\bibinfo {volume} {28}},\ \bibinfo
  {pages} {316} (\bibinfo {year} {1972})}\BibitemShut {NoStop}%
\bibitem [{\citenamefont {Pakvasa}(2000)}]{Pakvasa:1999ta}%
  \BibitemOpen
  \bibfield  {author} {\bibinfo {author} {\bibfnamefont {S.}~\bibnamefont
  {Pakvasa}},\ }\href {\doibase 10.1063/1.1336244} {\bibfield  {journal}
  {\bibinfo  {journal} {AIP Conf. Proc.}\ }\textbf {\bibinfo {volume} {542}},\
  \bibinfo {pages} {99} (\bibinfo {year} {2000})},\ \Eprint
  {http://arxiv.org/abs/hep-ph/0004077} {arXiv:hep-ph/0004077} \BibitemShut
  {NoStop}%
\bibitem [{\citenamefont {Acker}\ and\ \citenamefont
  {Pakvasa}(1994)}]{Acker:1993sz}%
  \BibitemOpen
  \bibfield  {author} {\bibinfo {author} {\bibfnamefont {A.}~\bibnamefont
  {Acker}}\ and\ \bibinfo {author} {\bibfnamefont {S.}~\bibnamefont
  {Pakvasa}},\ }\href {\doibase 10.1016/0370-2693(94)90663-7} {\bibfield
  {journal} {\bibinfo  {journal} {Phys. Lett. B}\ }\textbf {\bibinfo {volume}
  {320}},\ \bibinfo {pages} {320} (\bibinfo {year} {1994})},\ \Eprint
  {http://arxiv.org/abs/hep-ph/9310207} {arXiv:hep-ph/9310207} \BibitemShut
  {NoStop}%
\bibitem [{\citenamefont {Choubey}\ \emph {et~al.}(2000)\citenamefont
  {Choubey}, \citenamefont {Goswami},\ and\ \citenamefont
  {Majumdar}}]{Choubey:2000an}%
  \BibitemOpen
  \bibfield  {author} {\bibinfo {author} {\bibfnamefont {S.}~\bibnamefont
  {Choubey}}, \bibinfo {author} {\bibfnamefont {S.}~\bibnamefont {Goswami}}, \
  and\ \bibinfo {author} {\bibfnamefont {D.}~\bibnamefont {Majumdar}},\ }\href
  {\doibase 10.1016/S0370-2693(00)00608-0} {\bibfield  {journal} {\bibinfo
  {journal} {Phys. Lett. B}\ }\textbf {\bibinfo {volume} {484}},\ \bibinfo
  {pages} {73} (\bibinfo {year} {2000})},\ \Eprint
  {http://arxiv.org/abs/hep-ph/0004193} {arXiv:hep-ph/0004193} \BibitemShut
  {NoStop}%
\bibitem [{\citenamefont {Bandyopadhyay}\ \emph {et~al.}(2001)\citenamefont
  {Bandyopadhyay}, \citenamefont {Choubey},\ and\ \citenamefont
  {Goswami}}]{Bandyopadhyay:2001ct}%
  \BibitemOpen
  \bibfield  {author} {\bibinfo {author} {\bibfnamefont {A.}~\bibnamefont
  {Bandyopadhyay}}, \bibinfo {author} {\bibfnamefont {S.}~\bibnamefont
  {Choubey}}, \ and\ \bibinfo {author} {\bibfnamefont {S.}~\bibnamefont
  {Goswami}},\ }\href {\doibase 10.1103/PhysRevD.63.113019} {\bibfield
  {journal} {\bibinfo  {journal} {Phys. Rev. D}\ }\textbf {\bibinfo {volume}
  {63}},\ \bibinfo {pages} {113019} (\bibinfo {year} {2001})},\ \Eprint
  {http://arxiv.org/abs/hep-ph/0101273} {arXiv:hep-ph/0101273} \BibitemShut
  {NoStop}%
\bibitem [{\citenamefont {Joshipura}\ \emph {et~al.}(2002)\citenamefont
  {Joshipura}, \citenamefont {Masso},\ and\ \citenamefont
  {Mohanty}}]{Joshipura:2002fb}%
  \BibitemOpen
  \bibfield  {author} {\bibinfo {author} {\bibfnamefont {A.~S.}\ \bibnamefont
  {Joshipura}}, \bibinfo {author} {\bibfnamefont {E.}~\bibnamefont {Masso}}, \
  and\ \bibinfo {author} {\bibfnamefont {S.}~\bibnamefont {Mohanty}},\ }\href
  {\doibase 10.1103/PhysRevD.66.113008} {\bibfield  {journal} {\bibinfo
  {journal} {Phys. Rev. D}\ }\textbf {\bibinfo {volume} {66}},\ \bibinfo
  {pages} {113008} (\bibinfo {year} {2002})},\ \Eprint
  {http://arxiv.org/abs/hep-ph/0203181} {arXiv:hep-ph/0203181} \BibitemShut
  {NoStop}%
\bibitem [{\citenamefont {Bandyopadhyay}\ \emph {et~al.}(2003)\citenamefont
  {Bandyopadhyay}, \citenamefont {Choubey},\ and\ \citenamefont
  {Goswami}}]{Bandyopadhyay:2002qg}%
  \BibitemOpen
  \bibfield  {author} {\bibinfo {author} {\bibfnamefont {A.}~\bibnamefont
  {Bandyopadhyay}}, \bibinfo {author} {\bibfnamefont {S.}~\bibnamefont
  {Choubey}}, \ and\ \bibinfo {author} {\bibfnamefont {S.}~\bibnamefont
  {Goswami}},\ }\href {\doibase 10.1016/S0370-2693(03)00044-3} {\bibfield
  {journal} {\bibinfo  {journal} {Phys. Lett. B}\ }\textbf {\bibinfo {volume}
  {555}},\ \bibinfo {pages} {33} (\bibinfo {year} {2003})},\ \Eprint
  {http://arxiv.org/abs/hep-ph/0204173} {arXiv:hep-ph/0204173} \BibitemShut
  {NoStop}%
\bibitem [{\citenamefont {Berryman}\ \emph {et~al.}(2015)\citenamefont
  {Berryman}, \citenamefont {de~Gouvea},\ and\ \citenamefont
  {Hernandez}}]{Berryman:2014qha}%
  \BibitemOpen
  \bibfield  {author} {\bibinfo {author} {\bibfnamefont {J.~M.}\ \bibnamefont
  {Berryman}}, \bibinfo {author} {\bibfnamefont {A.}~\bibnamefont {de~Gouvea}},
  \ and\ \bibinfo {author} {\bibfnamefont {D.}~\bibnamefont {Hernandez}},\
  }\href {\doibase 10.1103/PhysRevD.92.073003} {\bibfield  {journal} {\bibinfo
  {journal} {Phys. Rev. D}\ }\textbf {\bibinfo {volume} {92}},\ \bibinfo
  {pages} {073003} (\bibinfo {year} {2015})},\ \Eprint
  {http://arxiv.org/abs/1411.0308} {arXiv:1411.0308 [hep-ph]} \BibitemShut
  {NoStop}%
\bibitem [{\citenamefont {Frieman}\ \emph {et~al.}(1988)\citenamefont
  {Frieman}, \citenamefont {Haber},\ and\ \citenamefont
  {Freese}}]{Frieman:1987as}%
  \BibitemOpen
  \bibfield  {author} {\bibinfo {author} {\bibfnamefont {J.~A.}\ \bibnamefont
  {Frieman}}, \bibinfo {author} {\bibfnamefont {H.~E.}\ \bibnamefont {Haber}},
  \ and\ \bibinfo {author} {\bibfnamefont {K.}~\bibnamefont {Freese}},\ }\href
  {\doibase 10.1016/0370-2693(88)91120-3} {\bibfield  {journal} {\bibinfo
  {journal} {Phys. Lett. B}\ }\textbf {\bibinfo {volume} {200}},\ \bibinfo
  {pages} {115} (\bibinfo {year} {1988})}\BibitemShut {NoStop}%
\bibitem [{\citenamefont {LoSecco}(1998)}]{LoSecco:1998cd}%
  \BibitemOpen
  \bibfield  {author} {\bibinfo {author} {\bibfnamefont {J.~M.}\ \bibnamefont
  {LoSecco}},\ }\href@noop {} {\  (\bibinfo {year} {1998})},\ \Eprint
  {http://arxiv.org/abs/hep-ph/9809499} {arXiv:hep-ph/9809499} \BibitemShut
  {NoStop}%
\bibitem [{\citenamefont {Barger}\ \emph {et~al.}(1999)\citenamefont {Barger},
  \citenamefont {Learned}, \citenamefont {Pakvasa},\ and\ \citenamefont
  {Weiler}}]{Barger:1998xk}%
  \BibitemOpen
  \bibfield  {author} {\bibinfo {author} {\bibfnamefont {V.~D.}\ \bibnamefont
  {Barger}}, \bibinfo {author} {\bibfnamefont {J.~G.}\ \bibnamefont {Learned}},
  \bibinfo {author} {\bibfnamefont {S.}~\bibnamefont {Pakvasa}}, \ and\
  \bibinfo {author} {\bibfnamefont {T.~J.}\ \bibnamefont {Weiler}},\ }\href
  {\doibase 10.1103/PhysRevLett.82.2640} {\bibfield  {journal} {\bibinfo
  {journal} {Phys. Rev. Lett.}\ }\textbf {\bibinfo {volume} {82}},\ \bibinfo
  {pages} {2640} (\bibinfo {year} {1999})},\ \Eprint
  {http://arxiv.org/abs/astro-ph/9810121} {arXiv:astro-ph/9810121} \BibitemShut
  {NoStop}%
\bibitem [{\citenamefont {Lipari}\ and\ \citenamefont
  {Lusignoli}(1999)}]{Lipari:1999vh}%
  \BibitemOpen
  \bibfield  {author} {\bibinfo {author} {\bibfnamefont {P.}~\bibnamefont
  {Lipari}}\ and\ \bibinfo {author} {\bibfnamefont {M.}~\bibnamefont
  {Lusignoli}},\ }\href {\doibase 10.1103/PhysRevD.60.013003} {\bibfield
  {journal} {\bibinfo  {journal} {Phys. Rev. D}\ }\textbf {\bibinfo {volume}
  {60}},\ \bibinfo {pages} {013003} (\bibinfo {year} {1999})},\ \Eprint
  {http://arxiv.org/abs/hep-ph/9901350} {arXiv:hep-ph/9901350} \BibitemShut
  {NoStop}%
\bibitem [{\citenamefont {Ashie}\ \emph {et~al.}(2004)\citenamefont {Ashie}
  \emph {et~al.}}]{Super-Kamiokande:2004orf}%
  \BibitemOpen
  \bibfield  {author} {\bibinfo {author} {\bibfnamefont {Y.}~\bibnamefont
  {Ashie}} \emph {et~al.} (\bibinfo {collaboration} {Super-Kamiokande}),\
  }\href {\doibase 10.1103/PhysRevLett.93.101801} {\bibfield  {journal}
  {\bibinfo  {journal} {Phys. Rev. Lett.}\ }\textbf {\bibinfo {volume} {93}},\
  \bibinfo {pages} {101801} (\bibinfo {year} {2004})},\ \Eprint
  {http://arxiv.org/abs/hep-ex/0404034} {arXiv:hep-ex/0404034} \BibitemShut
  {NoStop}%
\bibitem [{\citenamefont {Choubey}\ and\ \citenamefont
  {Goswami}(2000)}]{Choubey:1999ir}%
  \BibitemOpen
  \bibfield  {author} {\bibinfo {author} {\bibfnamefont {S.}~\bibnamefont
  {Choubey}}\ and\ \bibinfo {author} {\bibfnamefont {S.}~\bibnamefont
  {Goswami}},\ }\href {\doibase 10.1016/S0927-6505(00)00106-7} {\bibfield
  {journal} {\bibinfo  {journal} {Astropart. Phys.}\ }\textbf {\bibinfo
  {volume} {14}},\ \bibinfo {pages} {67} (\bibinfo {year} {2000})},\ \Eprint
  {http://arxiv.org/abs/hep-ph/9904257} {arXiv:hep-ph/9904257} \BibitemShut
  {NoStop}%
\bibitem [{\citenamefont {Gonzalez-Garcia}\ and\ \citenamefont
  {Maltoni}(2008)}]{Gonzalez-Garcia:2008mgl}%
  \BibitemOpen
  \bibfield  {author} {\bibinfo {author} {\bibfnamefont {M.~C.}\ \bibnamefont
  {Gonzalez-Garcia}}\ and\ \bibinfo {author} {\bibfnamefont {M.}~\bibnamefont
  {Maltoni}},\ }\href {\doibase 10.1016/j.physletb.2008.04.041} {\bibfield
  {journal} {\bibinfo  {journal} {Phys. Lett. B}\ }\textbf {\bibinfo {volume}
  {663}},\ \bibinfo {pages} {405} (\bibinfo {year} {2008})},\ \Eprint
  {http://arxiv.org/abs/0802.3699} {arXiv:0802.3699 [hep-ph]} \BibitemShut
  {NoStop}%
\bibitem [{\citenamefont {Gomes}\ \emph {et~al.}(2015)\citenamefont {Gomes},
  \citenamefont {Gomes},\ and\ \citenamefont {Peres}}]{Gomes:2014yua}%
  \BibitemOpen
  \bibfield  {author} {\bibinfo {author} {\bibfnamefont {R.~A.}\ \bibnamefont
  {Gomes}}, \bibinfo {author} {\bibfnamefont {A.~L.~G.}\ \bibnamefont {Gomes}},
  \ and\ \bibinfo {author} {\bibfnamefont {O.~L.~G.}\ \bibnamefont {Peres}},\
  }\href {\doibase 10.1016/j.physletb.2014.12.014} {\bibfield  {journal}
  {\bibinfo  {journal} {Phys. Lett. B}\ }\textbf {\bibinfo {volume} {740}},\
  \bibinfo {pages} {345} (\bibinfo {year} {2015})},\ \Eprint
  {http://arxiv.org/abs/1407.5640} {arXiv:1407.5640 [hep-ph]} \BibitemShut
  {NoStop}%
\bibitem [{\citenamefont {Choubey}\ \emph
  {et~al.}(2018{\natexlab{a}})\citenamefont {Choubey}, \citenamefont {Dutta},\
  and\ \citenamefont {Pramanik}}]{Choubey:2018cfz}%
  \BibitemOpen
  \bibfield  {author} {\bibinfo {author} {\bibfnamefont {S.}~\bibnamefont
  {Choubey}}, \bibinfo {author} {\bibfnamefont {D.}~\bibnamefont {Dutta}}, \
  and\ \bibinfo {author} {\bibfnamefont {D.}~\bibnamefont {Pramanik}},\ }\href
  {\doibase 10.1007/JHEP08(2018)141} {\bibfield  {journal} {\bibinfo  {journal}
  {JHEP}\ }\textbf {\bibinfo {volume} {08}},\ \bibinfo {pages} {141} (\bibinfo
  {year} {2018}{\natexlab{a}})},\ \Eprint {http://arxiv.org/abs/1805.01848}
  {arXiv:1805.01848 [hep-ph]} \BibitemShut {NoStop}%
\bibitem [{\citenamefont {Ternes}\ and\ \citenamefont
  {Pagliaroli}(2024)}]{Ternes:2024qui}%
  \BibitemOpen
  \bibfield  {author} {\bibinfo {author} {\bibfnamefont {C.~A.}\ \bibnamefont
  {Ternes}}\ and\ \bibinfo {author} {\bibfnamefont {G.}~\bibnamefont
  {Pagliaroli}},\ }\href@noop {} {\  (\bibinfo {year} {2024})},\ \Eprint
  {http://arxiv.org/abs/2401.14316} {arXiv:2401.14316 [hep-ph]} \BibitemShut
  {NoStop}%
\bibitem [{\citenamefont {Denton}\ and\ \citenamefont
  {Tamborra}(2018)}]{Denton:2018aml}%
  \BibitemOpen
  \bibfield  {author} {\bibinfo {author} {\bibfnamefont {P.~B.}\ \bibnamefont
  {Denton}}\ and\ \bibinfo {author} {\bibfnamefont {I.}~\bibnamefont
  {Tamborra}},\ }\href {\doibase 10.1103/PhysRevLett.121.121802} {\bibfield
  {journal} {\bibinfo  {journal} {Phys. Rev. Lett.}\ }\textbf {\bibinfo
  {volume} {121}},\ \bibinfo {pages} {121802} (\bibinfo {year} {2018})},\
  \Eprint {http://arxiv.org/abs/1805.05950} {arXiv:1805.05950 [hep-ph]}
  \BibitemShut {NoStop}%
\bibitem [{\citenamefont {Gr\"onroos}\ \emph {et~al.}(2024)\citenamefont
  {Gr\"onroos}, \citenamefont {Ohlsson},\ and\ \citenamefont
  {Vihonen}}]{Gronroos:2024jbs}%
  \BibitemOpen
  \bibfield  {author} {\bibinfo {author} {\bibfnamefont {J.}~\bibnamefont
  {Gr\"onroos}}, \bibinfo {author} {\bibfnamefont {T.}~\bibnamefont {Ohlsson}},
  \ and\ \bibinfo {author} {\bibfnamefont {S.}~\bibnamefont {Vihonen}},\
  }\href@noop {} {\  (\bibinfo {year} {2024})},\ \Eprint
  {http://arxiv.org/abs/2401.16864} {arXiv:2401.16864 [hep-ph]} \BibitemShut
  {NoStop}%
\bibitem [{\citenamefont {Choubey}\ \emph
  {et~al.}(2018{\natexlab{b}})\citenamefont {Choubey}, \citenamefont
  {Goswami},\ and\ \citenamefont {Pramanik}}]{Choubey:2017dyu}%
  \BibitemOpen
  \bibfield  {author} {\bibinfo {author} {\bibfnamefont {S.}~\bibnamefont
  {Choubey}}, \bibinfo {author} {\bibfnamefont {S.}~\bibnamefont {Goswami}}, \
  and\ \bibinfo {author} {\bibfnamefont {D.}~\bibnamefont {Pramanik}},\ }\href
  {\doibase 10.1007/JHEP02(2018)055} {\bibfield  {journal} {\bibinfo  {journal}
  {JHEP}\ }\textbf {\bibinfo {volume} {02}},\ \bibinfo {pages} {055} (\bibinfo
  {year} {2018}{\natexlab{b}})},\ \Eprint {http://arxiv.org/abs/1705.05820}
  {arXiv:1705.05820 [hep-ph]} \BibitemShut {NoStop}%
\bibitem [{\citenamefont {Ghoshal}\ \emph {et~al.}(2021)\citenamefont
  {Ghoshal}, \citenamefont {Giarnetti},\ and\ \citenamefont
  {Meloni}}]{Ghoshal:2020hyo}%
  \BibitemOpen
  \bibfield  {author} {\bibinfo {author} {\bibfnamefont {A.}~\bibnamefont
  {Ghoshal}}, \bibinfo {author} {\bibfnamefont {A.}~\bibnamefont {Giarnetti}},
  \ and\ \bibinfo {author} {\bibfnamefont {D.}~\bibnamefont {Meloni}},\ }\href
  {\doibase 10.1088/1361-6471/abdfab} {\bibfield  {journal} {\bibinfo
  {journal} {J. Phys. G}\ }\textbf {\bibinfo {volume} {48}},\ \bibinfo {pages}
  {055004} (\bibinfo {year} {2021})},\ \Eprint
  {http://arxiv.org/abs/2003.09012} {arXiv:2003.09012 [hep-ph]} \BibitemShut
  {NoStop}%
\bibitem [{\citenamefont {Tang}\ \emph {et~al.}(2019)\citenamefont {Tang},
  \citenamefont {Wang},\ and\ \citenamefont {Zhang}}]{Tang:2018rer}%
  \BibitemOpen
  \bibfield  {author} {\bibinfo {author} {\bibfnamefont {J.}~\bibnamefont
  {Tang}}, \bibinfo {author} {\bibfnamefont {T.-C.}\ \bibnamefont {Wang}}, \
  and\ \bibinfo {author} {\bibfnamefont {Y.}~\bibnamefont {Zhang}},\ }\href
  {\doibase 10.1007/JHEP04(2019)004} {\bibfield  {journal} {\bibinfo  {journal}
  {JHEP}\ }\textbf {\bibinfo {volume} {04}},\ \bibinfo {pages} {004} (\bibinfo
  {year} {2019})},\ \Eprint {http://arxiv.org/abs/1811.05623} {arXiv:1811.05623
  [hep-ph]} \BibitemShut {NoStop}%
\bibitem [{\citenamefont {Choubey}\ \emph {et~al.}(2021)\citenamefont
  {Choubey}, \citenamefont {Ghosh}, \citenamefont {Kempe},\ and\ \citenamefont
  {Ohlsson}}]{Choubey:2020dhw}%
  \BibitemOpen
  \bibfield  {author} {\bibinfo {author} {\bibfnamefont {S.}~\bibnamefont
  {Choubey}}, \bibinfo {author} {\bibfnamefont {M.}~\bibnamefont {Ghosh}},
  \bibinfo {author} {\bibfnamefont {D.}~\bibnamefont {Kempe}}, \ and\ \bibinfo
  {author} {\bibfnamefont {T.}~\bibnamefont {Ohlsson}},\ }\href {\doibase
  10.1007/JHEP05(2021)133} {\bibfield  {journal} {\bibinfo  {journal} {JHEP}\
  }\textbf {\bibinfo {volume} {05}},\ \bibinfo {pages} {133} (\bibinfo {year}
  {2021})},\ \Eprint {http://arxiv.org/abs/2010.16334} {arXiv:2010.16334
  [hep-ph]} \BibitemShut {NoStop}%
\bibitem [{\citenamefont {Chakraborty}\ \emph {et~al.}(2021)\citenamefont
  {Chakraborty}, \citenamefont {Dutta}, \citenamefont {Goswami},\ and\
  \citenamefont {Pramanik}}]{Chakraborty:2020cfu}%
  \BibitemOpen
  \bibfield  {author} {\bibinfo {author} {\bibfnamefont {K.}~\bibnamefont
  {Chakraborty}}, \bibinfo {author} {\bibfnamefont {D.}~\bibnamefont {Dutta}},
  \bibinfo {author} {\bibfnamefont {S.}~\bibnamefont {Goswami}}, \ and\
  \bibinfo {author} {\bibfnamefont {D.}~\bibnamefont {Pramanik}},\ }\href
  {\doibase 10.1007/JHEP08(2021)136} {\bibfield  {journal} {\bibinfo  {journal}
  {JHEP}\ }\textbf {\bibinfo {volume} {08}},\ \bibinfo {pages} {136} (\bibinfo
  {year} {2021})},\ \Eprint {http://arxiv.org/abs/2012.04958} {arXiv:2012.04958
  [hep-ph]} \BibitemShut {NoStop}%
\bibitem [{\citenamefont {Abrah\~ao}\ \emph {et~al.}(2015)\citenamefont
  {Abrah\~ao}, \citenamefont {Minakata}, \citenamefont {Nunokawa},\ and\
  \citenamefont {Quiroga}}]{Abrahao:2015rba}%
  \BibitemOpen
  \bibfield  {author} {\bibinfo {author} {\bibfnamefont {T.}~\bibnamefont
  {Abrah\~ao}}, \bibinfo {author} {\bibfnamefont {H.}~\bibnamefont {Minakata}},
  \bibinfo {author} {\bibfnamefont {H.}~\bibnamefont {Nunokawa}}, \ and\
  \bibinfo {author} {\bibfnamefont {A.~A.}\ \bibnamefont {Quiroga}},\ }\href
  {\doibase 10.1007/JHEP11(2015)001} {\bibfield  {journal} {\bibinfo  {journal}
  {JHEP}\ }\textbf {\bibinfo {volume} {11}},\ \bibinfo {pages} {001} (\bibinfo
  {year} {2015})},\ \Eprint {http://arxiv.org/abs/1506.02314} {arXiv:1506.02314
  [hep-ph]} \BibitemShut {NoStop}%
\bibitem [{\citenamefont {Huang}\ and\ \citenamefont
  {Zhou}(2019)}]{Huang:2018nxj}%
  \BibitemOpen
  \bibfield  {author} {\bibinfo {author} {\bibfnamefont {G.-Y.}\ \bibnamefont
  {Huang}}\ and\ \bibinfo {author} {\bibfnamefont {S.}~\bibnamefont {Zhou}},\
  }\href {\doibase 10.1088/1475-7516/2019/02/024} {\bibfield  {journal}
  {\bibinfo  {journal} {JCAP}\ }\textbf {\bibinfo {volume} {02}},\ \bibinfo
  {pages} {024} (\bibinfo {year} {2019})},\ \Eprint
  {http://arxiv.org/abs/1810.03877} {arXiv:1810.03877 [hep-ph]} \BibitemShut
  {NoStop}%
\bibitem [{\citenamefont {Mart\'\i{}nez-Mirav\'e}\ \emph
  {et~al.}(2024)\citenamefont {Mart\'\i{}nez-Mirav\'e}, \citenamefont
  {Tamborra},\ and\ \citenamefont {T\'ortola}}]{Martinez-Mirave:2024hfd}%
  \BibitemOpen
  \bibfield  {author} {\bibinfo {author} {\bibfnamefont {P.}~\bibnamefont
  {Mart\'\i{}nez-Mirav\'e}}, \bibinfo {author} {\bibfnamefont {I.}~\bibnamefont
  {Tamborra}}, \ and\ \bibinfo {author} {\bibfnamefont {M.}~\bibnamefont
  {T\'ortola}},\ }\href@noop {} {\  (\bibinfo {year} {2024})},\ \Eprint
  {http://arxiv.org/abs/2402.00116} {arXiv:2402.00116 [astro-ph.HE]}
  \BibitemShut {NoStop}%
\bibitem [{\citenamefont {Gago}\ \emph {et~al.}(2017)\citenamefont {Gago},
  \citenamefont {Gomes}, \citenamefont {Gomes}, \citenamefont {Jones-Perez},\
  and\ \citenamefont {Peres}}]{Gago:2017zzy}%
  \BibitemOpen
  \bibfield  {author} {\bibinfo {author} {\bibfnamefont {A.~M.}\ \bibnamefont
  {Gago}}, \bibinfo {author} {\bibfnamefont {R.~A.}\ \bibnamefont {Gomes}},
  \bibinfo {author} {\bibfnamefont {A.~L.~G.}\ \bibnamefont {Gomes}}, \bibinfo
  {author} {\bibfnamefont {J.}~\bibnamefont {Jones-Perez}}, \ and\ \bibinfo
  {author} {\bibfnamefont {O.~L.~G.}\ \bibnamefont {Peres}},\ }\href {\doibase
  10.1007/JHEP11(2017)022} {\bibfield  {journal} {\bibinfo  {journal} {JHEP}\
  }\textbf {\bibinfo {volume} {11}},\ \bibinfo {pages} {022} (\bibinfo {year}
  {2017})},\ \Eprint {http://arxiv.org/abs/1705.03074} {arXiv:1705.03074
  [hep-ph]} \BibitemShut {NoStop}%
\bibitem [{\citenamefont {Coloma}\ and\ \citenamefont
  {Peres}(2017)}]{Coloma:2017zpg}%
  \BibitemOpen
  \bibfield  {author} {\bibinfo {author} {\bibfnamefont {P.}~\bibnamefont
  {Coloma}}\ and\ \bibinfo {author} {\bibfnamefont {O.~L.~G.}\ \bibnamefont
  {Peres}},\ }\href@noop {} {\  (\bibinfo {year} {2017})},\ \Eprint
  {http://arxiv.org/abs/1705.03599} {arXiv:1705.03599 [hep-ph]} \BibitemShut
  {NoStop}%
\bibitem [{\citenamefont {Ascencio-Sosa}\ \emph {et~al.}(2018)\citenamefont
  {Ascencio-Sosa}, \citenamefont {Calatayud-Cadenillas}, \citenamefont {Gago},\
  and\ \citenamefont {Jones-P\'erez}}]{Ascencio-Sosa:2018lbk}%
  \BibitemOpen
  \bibfield  {author} {\bibinfo {author} {\bibfnamefont {M.~V.}\ \bibnamefont
  {Ascencio-Sosa}}, \bibinfo {author} {\bibfnamefont {A.~M.}\ \bibnamefont
  {Calatayud-Cadenillas}}, \bibinfo {author} {\bibfnamefont {A.~M.}\
  \bibnamefont {Gago}}, \ and\ \bibinfo {author} {\bibfnamefont
  {J.}~\bibnamefont {Jones-P\'erez}},\ }\href {\doibase
  10.1140/epjc/s10052-018-6276-0} {\bibfield  {journal} {\bibinfo  {journal}
  {Eur. Phys. J. C}\ }\textbf {\bibinfo {volume} {78}},\ \bibinfo {pages} {809}
  (\bibinfo {year} {2018})},\ \Eprint {http://arxiv.org/abs/1805.03279}
  {arXiv:1805.03279 [hep-ph]} \BibitemShut {NoStop}%
\bibitem [{\citenamefont {Barger}\ \emph {et~al.}(2002)\citenamefont {Barger},
  \citenamefont {Marfatia},\ and\ \citenamefont {Whisnant}}]{Barger:2001yr}%
  \BibitemOpen
  \bibfield  {author} {\bibinfo {author} {\bibfnamefont {V.}~\bibnamefont
  {Barger}}, \bibinfo {author} {\bibfnamefont {D.}~\bibnamefont {Marfatia}}, \
  and\ \bibinfo {author} {\bibfnamefont {K.}~\bibnamefont {Whisnant}},\ }\href
  {\doibase 10.1103/PhysRevD.65.073023} {\bibfield  {journal} {\bibinfo
  {journal} {Phys. Rev. D}\ }\textbf {\bibinfo {volume} {65}},\ \bibinfo
  {pages} {073023} (\bibinfo {year} {2002})},\ \Eprint
  {http://arxiv.org/abs/hep-ph/0112119} {arXiv:hep-ph/0112119} \BibitemShut
  {NoStop}%
\bibitem [{\citenamefont {Abi}\ \emph {et~al.}(2021{\natexlab{b}})\citenamefont
  {Abi} \emph {et~al.}}]{DUNE:2021cuw}%
  \BibitemOpen
  \bibfield  {author} {\bibinfo {author} {\bibfnamefont {B.}~\bibnamefont
  {Abi}} \emph {et~al.} (\bibinfo {collaboration} {DUNE}),\ }\href@noop {} {\
  (\bibinfo {year} {2021}{\natexlab{b}})},\ \Eprint
  {http://arxiv.org/abs/2103.04797} {arXiv:2103.04797 [hep-ex]} \BibitemShut
  {NoStop}%
\bibitem [{\citenamefont {De~Romeri}\ \emph {et~al.}(2016)\citenamefont
  {De~Romeri}, \citenamefont {Fernandez-Martinez},\ and\ \citenamefont
  {Sorel}}]{DeRomeri:2016qwo}%
  \BibitemOpen
  \bibfield  {author} {\bibinfo {author} {\bibfnamefont {V.}~\bibnamefont
  {De~Romeri}}, \bibinfo {author} {\bibfnamefont {E.}~\bibnamefont
  {Fernandez-Martinez}}, \ and\ \bibinfo {author} {\bibfnamefont
  {M.}~\bibnamefont {Sorel}},\ }\href {\doibase 10.1007/JHEP09(2016)030}
  {\bibfield  {journal} {\bibinfo  {journal} {JHEP}\ }\textbf {\bibinfo
  {volume} {09}},\ \bibinfo {pages} {030} (\bibinfo {year} {2016})},\ \Eprint
  {http://arxiv.org/abs/1607.00293} {arXiv:1607.00293 [hep-ph]} \BibitemShut
  {NoStop}%
\bibitem [{\citenamefont {Abe}\ \emph {et~al.}(2018)\citenamefont {Abe} \emph
  {et~al.}}]{Hyper-Kamiokande:2016srs}%
  \BibitemOpen
  \bibfield  {author} {\bibinfo {author} {\bibfnamefont {K.}~\bibnamefont
  {Abe}} \emph {et~al.} (\bibinfo {collaboration} {Hyper-Kamiokande}),\ }\href
  {\doibase 10.1093/ptep/pty044} {\bibfield  {journal} {\bibinfo  {journal}
  {PTEP}\ }\textbf {\bibinfo {volume} {2018}},\ \bibinfo {pages} {063C01}
  (\bibinfo {year} {2018})},\ \Eprint {http://arxiv.org/abs/1611.06118}
  {arXiv:1611.06118 [hep-ex]} \BibitemShut {NoStop}%
\bibitem [{\citenamefont {Huber}\ \emph {et~al.}(2005)\citenamefont {Huber},
  \citenamefont {Lindner},\ and\ \citenamefont {Winter}}]{Huber:2004ka}%
  \BibitemOpen
  \bibfield  {author} {\bibinfo {author} {\bibfnamefont {P.}~\bibnamefont
  {Huber}}, \bibinfo {author} {\bibfnamefont {M.}~\bibnamefont {Lindner}}, \
  and\ \bibinfo {author} {\bibfnamefont {W.}~\bibnamefont {Winter}},\ }\href
  {\doibase 10.1016/j.cpc.2005.01.003} {\bibfield  {journal} {\bibinfo
  {journal} {Comput. Phys. Commun.}\ }\textbf {\bibinfo {volume} {167}},\
  \bibinfo {pages} {195} (\bibinfo {year} {2005})},\ \Eprint
  {http://arxiv.org/abs/hep-ph/0407333} {arXiv:hep-ph/0407333} \BibitemShut
  {NoStop}%
\bibitem [{\citenamefont {Esteban}\ \emph {et~al.}(2020)\citenamefont
  {Esteban}, \citenamefont {Gonzalez-Garcia}, \citenamefont {Maltoni},
  \citenamefont {Schwetz},\ and\ \citenamefont {Zhou}}]{Esteban:2020cvm}%
  \BibitemOpen
  \bibfield  {author} {\bibinfo {author} {\bibfnamefont {I.}~\bibnamefont
  {Esteban}}, \bibinfo {author} {\bibfnamefont {M.~C.}\ \bibnamefont
  {Gonzalez-Garcia}}, \bibinfo {author} {\bibfnamefont {M.}~\bibnamefont
  {Maltoni}}, \bibinfo {author} {\bibfnamefont {T.}~\bibnamefont {Schwetz}}, \
  and\ \bibinfo {author} {\bibfnamefont {A.}~\bibnamefont {Zhou}},\ }\href
  {\doibase 10.1007/JHEP09(2020)178} {\bibfield  {journal} {\bibinfo  {journal}
  {JHEP}\ }\textbf {\bibinfo {volume} {09}},\ \bibinfo {pages} {178} (\bibinfo
  {year} {2020})},\ \Eprint {http://arxiv.org/abs/2007.14792} {arXiv:2007.14792
  [hep-ph]} \BibitemShut {NoStop}%
\bibitem [{\citenamefont {Choubey}\ \emph
  {et~al.}(2018{\natexlab{c}})\citenamefont {Choubey}, \citenamefont {Goswami},
  \citenamefont {Gupta}, \citenamefont {Lakshmi},\ and\ \citenamefont
  {Thakore}}]{Choubey:2017eyg}%
  \BibitemOpen
  \bibfield  {author} {\bibinfo {author} {\bibfnamefont {S.}~\bibnamefont
  {Choubey}}, \bibinfo {author} {\bibfnamefont {S.}~\bibnamefont {Goswami}},
  \bibinfo {author} {\bibfnamefont {C.}~\bibnamefont {Gupta}}, \bibinfo
  {author} {\bibfnamefont {S.~M.}\ \bibnamefont {Lakshmi}}, \ and\ \bibinfo
  {author} {\bibfnamefont {T.}~\bibnamefont {Thakore}},\ }\href {\doibase
  10.1103/PhysRevD.97.033005} {\bibfield  {journal} {\bibinfo  {journal} {Phys.
  Rev. D}\ }\textbf {\bibinfo {volume} {97}},\ \bibinfo {pages} {033005}
  (\bibinfo {year} {2018}{\natexlab{c}})},\ \Eprint
  {http://arxiv.org/abs/1709.10376} {arXiv:1709.10376 [hep-ph]} \BibitemShut
  {NoStop}%
\bibitem [{\citenamefont {Chacko}\ \emph {et~al.}(2020)\citenamefont {Chacko},
  \citenamefont {Dev}, \citenamefont {Du}, \citenamefont {Poulin},\ and\
  \citenamefont {Tsai}}]{Chacko:2019nej}%
  \BibitemOpen
  \bibfield  {author} {\bibinfo {author} {\bibfnamefont {Z.}~\bibnamefont
  {Chacko}}, \bibinfo {author} {\bibfnamefont {A.}~\bibnamefont {Dev}},
  \bibinfo {author} {\bibfnamefont {P.}~\bibnamefont {Du}}, \bibinfo {author}
  {\bibfnamefont {V.}~\bibnamefont {Poulin}}, \ and\ \bibinfo {author}
  {\bibfnamefont {Y.}~\bibnamefont {Tsai}},\ }\href {\doibase
  10.1007/JHEP04(2020)020} {\bibfield  {journal} {\bibinfo  {journal} {JHEP}\
  }\textbf {\bibinfo {volume} {04}},\ \bibinfo {pages} {020} (\bibinfo {year}
  {2020})},\ \Eprint {http://arxiv.org/abs/1909.05275} {arXiv:1909.05275
  [hep-ph]} \BibitemShut {NoStop}%
\bibitem [{\citenamefont {Escudero}\ \emph {et~al.}(2020)\citenamefont
  {Escudero}, \citenamefont {Lopez-Pavon}, \citenamefont {Rius},\ and\
  \citenamefont {Sandner}}]{Escudero:2020ped}%
  \BibitemOpen
  \bibfield  {author} {\bibinfo {author} {\bibfnamefont {M.}~\bibnamefont
  {Escudero}}, \bibinfo {author} {\bibfnamefont {J.}~\bibnamefont
  {Lopez-Pavon}}, \bibinfo {author} {\bibfnamefont {N.}~\bibnamefont {Rius}}, \
  and\ \bibinfo {author} {\bibfnamefont {S.}~\bibnamefont {Sandner}},\ }\href
  {\doibase 10.1007/JHEP12(2020)119} {\bibfield  {journal} {\bibinfo  {journal}
  {JHEP}\ }\textbf {\bibinfo {volume} {12}},\ \bibinfo {pages} {119} (\bibinfo
  {year} {2020})},\ \Eprint {http://arxiv.org/abs/2007.04994} {arXiv:2007.04994
  [hep-ph]} \BibitemShut {NoStop}%
\end{thebibliography}%
\bibliographystyle{apsrev4-1}

\captionsetup{font=normalsize}

\end{document}